\documentclass[11pt]{article}
\usepackage{axodraw}
\usepackage{epsfig}
\usepackage{amsfonts}
\usepackage{amsmath}
\usepackage{bbm}
\allowdisplaybreaks[1]

\input pix.sty

\newcommand{\Lii}{\,\mbox{Li}_\rmi{2}}
\newcommand{\Liii}{\,\mbox{Li}_\rmi{3}}

\newcommand{\ko}{k_0}
\newcommand{\km}{k_-}
\newcommand{\kp}{k_+}

\renewcommand{\eq}{eq.~}
\renewcommand{\eqs}{eqs.~}
\renewcommand{\se}{sec.~}

\renewcommand{\fig}{fig.~}

\newcommand{\tinymsbar}{{\overline{\mbox{\tiny\rm{MS}}}}}
\newcommand{\Lambdamsbar}{{\Lambda_\tinymsbar}}

\newcommand{\Nf}{N_{\rm f}}
\newcommand{\Nc}{N_{\rm c}}

\newcommand{\Tc}{T_{\rm c}}
\newcommand{\CF}{C_\rmii{F}}

\newcommand{\rmO}{{\mathcal{O}}}
\newcommand{\bmu}{\bar\mu}

\def\lsi{\raise0.3ex\hbox{$<$\kern-0.75em\raise-1.1ex\hbox{$\sim$}}}
\def\gsi{\raise0.3ex\hbox{$>$\kern-0.75em\raise-1.1ex\hbox{$\sim$}}}
\newcommand{\lsim}{\mathop{\lsi}}
\newcommand{\gsim}{\mathop{\gsi}}

\newcommand{\sign}{\mathop{\mbox{sign}}}
\newcommand{\nF}{n_\rmii{F}}
\newcommand{\nB}{n_\rmii{B}}
 \renewcommand{\nF}[1]{n_\rmii{F{#1}}}
 \renewcommand{\nB}[1]{n_\rmii{B{#1}}}

\newcommand{\rmii}[1]{{\mbox{\tiny\rm{#1}}}}

\newcommand{\im}{\mathop{\mbox{Im}}}

\newcommand{\Tint}[1]{{\hbox{$\sum$}\!\!\!\!\!\!\!\int\,}_{\!\!\!\!\raise-0.9ex\hbox{$\scriptstyle{#1}$}}}
\newcommand{\Tinti}[1]{{{\Sigma}\!\!\!\!\raise0.3ex\hbox{$\int$}_\rmii{${#1}$}}}

\newcommand{\bi}{\begin{itemize}}
\newcommand{\ei}{\end{itemize}}

\newcommand{\hide}[1]{ }

\def\TAsc(#1,#2)(#3,#4,#5)%
{\SetWidth{2.0}\CArc(#1,#2)(#3,#4,#5)\SetWidth{1.0}}
\def\Lwidth{3}

\def\TAgl(#1,#2)(#3,#4,#5){\SetWidth{2.0}\PhotonArc(#1,#2)(#3,#4,#5){\Lwidth}%
{6.283 #3 mul 360 div #4 #5 sub #4 #5 sub mul sqrt mul Tdensity mul}%
\SetWidth{1.0}}
\def\TLgl(#1,#2)(#3,#4){\SetWidth{2.0}\Photon(#1,#2)(#3,#4){\Lwidth}
{#1 #3 sub #1 #3 sub mul #2 #4 sub #2 #4 sub mul add sqrt Tdensity mul}%
\SetWidth{1.0}}

\renewcommand{\picc}[1]{\;\parbox[c]{60pt}{\begin{picture}(60,30)(0,0)
\SetWidth{1.0}\SetScale{1.3} #1 \end{picture}}\; }
\def\Lwidth{1.3}

\newcommand{\picv}[1]{\;\parbox[c]{80pt}{\begin{picture}(80,70)(0,-5)
\SetWidth{1.0}\SetScale{1.0} #1 \end{picture}}\; }
\def\Generic{\picv{%
 \CArc(40,30)(30,0,360)%
 \DashLine(-10,30)(10,30){5}%
 \DashLine(70,30)(90,30){5}%
 \DashLine(40,0)(40,60){5}%
 \Text(-15,30)[r]{$\scriptstyle \sigma_0, K$}%
 \Text(95,30)[l]{$\scriptstyle \sigma_0, K$}%
 \Text(60,60)[l]{$\scriptstyle \sigma_2, Q$}%
 \Text(27,30)[l]{$\scriptstyle \sigma_5, \;\, Q-P$}
 \Text(20,0)[r]{$\scriptstyle \sigma_4, P-K$}%
 \Text(20,60)[r]{$\scriptstyle \sigma_1, P$}%
 \Text(60,0)[l]{$\scriptstyle \sigma_3, Q-K$}%
}}
\makeatletter \@addtoreset{equation}{section} \makeatother
\renewcommand{\theequation}{\arabic{section}.\arabic{equation}}
\makeatletter
\renewcommand\section{\@startsection {section}{1}{\z@}%
                                   {-5.5ex \@plus -1ex \@minus -.2ex}
                                   {2.3ex \@plus.2ex}%
                                   {\normalfont\large\bfseries}}
\renewcommand\subsection{\@startsection{subsection}{2}{\z@}%
                                     {-3.25ex\@plus -1ex \@minus -.2ex}%
                                     {1.5ex \@plus .2ex}%
                                     {\normalfont\normalsize\bfseries}}
\renewcommand\thesection {\@arabic\c@section}
\renewcommand\thesubsection   {\thesection.\@arabic\c@subsection}
\renewcommand{\@seccntformat}[1]{%
\csname the#1\endcsname.\hspace{1.0em}}
\makeatother

\begin{document}

\flushbottom

\begin{titlepage}

\begin{flushright}
\vspace*{1cm}
\end{flushright}
\begin{centering}
\vfill

{\Large{\bf
 NLO thermal dilepton rate at non-zero momentum
}} 

\vspace{0.8cm}

M.~Laine 

\vspace{0.8cm}

{\em
Institute for Theoretical Physics, 
Albert Einstein Center, University of Bern, \\ 
Sidlerstrasse 5, CH-3012 Bern, Switzerland\\}


\vspace*{0.8cm}

\mbox{\bf Abstract}
 
\end{centering}

\vspace*{0.3cm}
 
\noindent
The vector channel spectral function and the dilepton production rate
from a QCD plasma at a temperature above a few hundred MeV are
evaluated up to next-to-leading order (NLO) including their
dependence on a non-zero momentum with respect to the heat bath. 
The invariant mass of the virtual photon is taken to be in the range
$\mathcal{K}^2 \sim (\pi T)^2 \sim (\mbox{1 GeV})^2$, generalizing
previous NLO results valid for $\mathcal{K}^2 \gg (\pi T)^2$. In the
opposite regime $0 < \mathcal{K}^2 \ll (\pi T)^2$ the loop expansion
breaks down, but agrees nevertheless in order of magnitude with
a previous result obtained through resummations. 
Ways to test the vector spectral function through comparisons with 
imaginary-time correlators measured on the lattice are discussed.

\vfill

 
\vspace*{1cm}
  
\noindent
October 2013

\vfill

\end{titlepage}

%
\section{Introduction}

Characteristics of ``hard probes'', produced within a thermal medium but 
immediately escaping it, are theoretically among the best ways to
learn about the properties of the medium. For a plasma made of 
strongly interacting particles at a temperature of a few hundred MeV, 
which has a spatial extent of some tens of fm, 
typical hard probes are particles only 
experiencing weak and electromagnetic interactions, such as photons
and leptons. Indeed the photon and dilepton production rates from 
a quark-gluon plasma have been studied in great detail
in the last three decades (cf.\ e.g.\ refs.~\cite{mlb,kg}). 

Even though hard probes behave as free particles once produced, 
their production mechanism is complicated, due to the strong
interactions felt by the quarks and gluons that form the plasma. Therefore, 
despite the long history, our knowledge of the differential production
rate as a function of temperature, baryon chemical potential, and
quark mass spectrum, remains
incomplete. In fact, in much of the parameter space, 
only the leading-order (LO) result in the strong coupling constant,
$\alpha_s$, is available. Given that $\alpha_s$ is not small 
at temperatures reached in practical heavy ion collision 
experiments, this could imply errors of up to 50\% or so. 
It would be desirable to find ways to reduce the uncertainty, 
and the current paper aims to play a role in this endeavour, 
by determining novel next-to-leading order (NLO) corrections to the 
production rate of a virtual photon subsequently decaying 
into a dilepton pair.   

Denoting  the four-momentum of the virtual photon, measured
in the rest frame of the heat bath, by 
$\mathcal{K} = (\ko,\vec{k})$, with $k \equiv |\vec{k}|$ and 
$\mathcal{K}^2 \equiv \ko^2 - k^2$,  
the corresponding production rate (encoded
in a {\em vector channel spectral function}) 
has been computed up to NLO, 
or $\rmO(\alpha_s)$, at 
$k=0$ both for massless
($m \ll \pi T$)~\cite{old1,old2,old3} and for heavy ($m \gg \pi T$)
quarks~\cite{nlo}. At $k\neq 0$ only 
its asymptotics at $\mathcal{K}^2 \gg (\pi T)^2$  has
been determined up to $\rmO(\alpha_s)$~\cite{sch}. 
The vacuum part of the vector spectral function
is known up to N$^4$LO, or $\rmO(\alpha_s^4)$~\cite{kit_ns,kit_si}.
The goal of the present study is to complete the $\rmO(\alpha_s)$
level for a general $k\neq 0$, $T\neq 0$, with 
$\mathcal{K}^2 \sim (\pi T)^2$.

In contrast, for $0 < \mathcal{K}^2 \ll (\pi T)^2$, multiple 
scatterings take place within a typical formation time of 
the virtual photon. Consequently the loop expansion 
breaks down and needs to be resummed even to obtain the correct LO result. 
(The celebrated Hard Thermal Loop resummation~\cite{bp} is not enough 
to render the loop expansion convergent on its own but a further 
all-orders resummation is needed, 
cf.\ ref.~\cite{bb} for a recent discussion.) 
For the dilepton rate a resummation has been 
implemented both for $k \neq 0$~\cite{agz} 
(according to ref.~\cite{cg} these results contain a slight error)
and for $k=0$~\cite{mr}, following previous work
on the production rate of on-shell photons~\cite{amy1,amy2,amy3}. 
Even though the current paper cannot serve as 
a crosscheck of these resummations, it is nevertheless
comforting that a conceptually much simpler procedure 
yields qualitatively similar results as long as $\mathcal{K}^2$ is non-zero.  

Of course, particularly for $0 < \mathcal{K}^2 \ll (\pi T)^2$, 
it may also be questioned whether the results
are quantitatively accurate even if systematically resummed, 
because the effective value of $\alpha_s$ may be substantial.  
With this motivation in mind there have been attempts at lattice 
estimates of the vector channel spectral function. 
For quenched QCD, measurements of imaginary-time correlators are 
approaching the continuum limit both for $k=0$~\cite{hengtong1} 
and $k\neq 0$~\cite{hengtong2}.
Even though current extractions of the spectral functions may
contain uncontrolled uncertainties~\cite{cond}, further 
progress will undoubtedly follow. Recent results also exist
for dynamical quarks~\cite{mainz,swansea}, however
in this case no continuum extrapolation has been carried out and 
systematic uncertainties are correspondingly larger. 
As will be discussed below a strict comparison of the perturbative
and lattice results is not possible at present because of 
missing ingredients on both sides; nevertheless, on a semi-quantitative
level a good agreement is found (for this it is essential that 
a continuum extrapolation has been carried out). 

We start by outlining the ingredients of an NLO computation
at $\mathcal{K}^2 \sim (\pi T)^2$ in \se\ref{se:setup}. 
The main results are given in \se\ref{se:results}, together
with comparisons with various limiting values. Numerical
results for the vector channel spectral function and dilepton
spectra are shown in \se\ref{se:spectra}, whereas 
in \se\ref{se:eucl} the corresponding imaginary-time correlators, 
measurable with lattice simulations, are discussed. 
Sec.~\ref{se:concl} offers a brief summary; two appendices 
collect various technical details related to the computation. 
(However the main computational ingredients, recently worked
out in refs.~\cite{master,rel}, will not be re-discussed here.)

%
\section{Setup}
\la{se:setup}

To leading order in 
$\alpha^{ }_{e} \equiv e^2/(4\pi)$~\cite{dilepton1,dilepton2,dilepton3} 
and $\alpha^{ }_w \equiv g_w^2/(4\pi)$,  
the production rate of 
$\mu^-\mu^+$ (or $e^-e^+$) pairs from a hot QCD medium, 
with a total four-momentum $\mathcal{K}$, 
can be expressed as
\ba
 && \hspace*{-1cm}
 \frac{{\rm d} N_{\mu^-\mu^+}}
   {{\rm d}^4 \mathcal{X} {\rm d}^4 \mathcal{K}} 
  \; \stackrel{4 m_\mu^2 \ll \mathcal{K}^2 \ll m_Z^2 }{=} \;   
 - \frac{ \nB{} (\ko)  } 
  {3 \pi^3 \mathcal{K}^2} 
 \biggl( \eta_{\mu\nu} -  \frac{\mathcal{K}_\mu\mathcal{K}_\nu}{\mathcal{K}^2}
 \biggr)
 \nn & \times & 
 \biggl\{ 
 \alpha_{e}^2 
 \biggl[\Bigl( 
 \sum_{i = 1}^{\Nf} Q_i^2\Bigr) \;  
 \rho_\rmii{V,NS}^{\mu\nu}(\mathcal{K}) + 
 \Bigl( \sum_{i = 1}^{\Nf} Q_i \Bigr)^2 \; 
 \rho_\rmii{V,SI}^{\mu\nu}(\mathcal{K}) 
 \biggr]
 \nn & & \; + 
 \frac{\alpha_{e}\alpha_{w} \mathcal{K}^2}{\mathcal{K}^2 - m_Z^2}
 \frac{1 - 4 s_w^2}{8(1-s_w^2)}
 \biggl[
 \Bigl( \sum_{i = 1}^{\Nf} Q_i C_{i,v} \Bigr) \;  
 \rho_\rmii{V,NS}^{\mu\nu}(\mathcal{K}) + 
 \Bigl( \sum_{i = 1}^{\Nf} Q_i \Bigr) 
 \Bigl( \sum_{i = 1}^{\Nf} C_{i,v} \Bigr) \; 
 \rho_\rmii{V,SI}^{\mu\nu}(\mathcal{K}) 
 \biggr]
 \nn & & \; + \,
 \frac{\alpha_{w}^2 \mathcal{K}^4}{(\mathcal{K}^2 - m_Z^2)^2}
 \frac{1+(1 - 4 s_w^2)^2}{256(1-s_w^2)^2}
 \biggl[
 \Bigl( \sum_{i = 1}^{\Nf} C_{i,v}^2 \Bigr) \;  
 \rho_\rmii{V,NS}^{\mu\nu}(\mathcal{K}) + 
 \Bigl( \sum_{i = 1}^{\Nf} C_{i,v} \Bigr)^2 \; 
 \rho_\rmii{V,SI}^{\mu\nu}(\mathcal{K})
 \nn & & \qquad  + \,
 \Bigl( \sum_{i = 1}^{\Nf} C_{i,a}^2 \Bigr) \;  
 \rho_\rmii{A,NS}^{\mu\nu}(\mathcal{K}) + 
 \Bigl( \sum_{i = 1}^{\Nf} C_{i,a}^{ } \Bigr)^2 \; 
 \rho_\rmii{A,SI}^{\mu\nu}(\mathcal{K})
 \biggr]
  \biggr\}
 \;, 
 \la{physics} 
\ea
where $\nB{}$ is the Bose distribution,
$s_w \equiv \sin \theta_w$ the weak mixing angle,  
$Q_{i} \in (\fr23,-\fr13)$ the electric charge of quark of flavour $i$  
in units of $e$, and $C_{i,v} \in (1-\fr83 s_w^2,-1+\fr43 s_w^2)$, 
$C_{i,a} \in (-1,1)$ parametrise the vector and axial neutral-current 
couplings of up and down-type quark flavours, respectively. 
By $\rho^{\mu\nu}_\rmii{V,NS}$ we denote
the spectral function corresponding to the vector current
in the ``non-singlet'' (NS) channel arising from 
quark-connected contractions; the ``singlet'' (SI) contributions
arise instead from disconnected quark contractions. 
Flavour-degenerate quark masses have been assumed for the 
reduction in \eq\nr{physics}. 
Setting the quark masses furthermore to zero, a Ward identity guarantees 
(in the absence of quark zero-mode contributions)
that the non-singlet axial current spectral function 
($\rho^{\mu\nu}_\rmii{A,NS}$) agrees with the vector one. 
Moreover the singlet channels are 
suppressed by $\alpha_s^3$~\cite{kit_si}.
Therefore we concentrate on the non-singlet
vector channel; because of current conservation
it can be expressed as 
\ba
 \rho_\rmii{NS}(\mathcal{K}) & \equiv & 
 \biggl( \eta_{\mu\nu} -  \frac{\mathcal{K}_\mu\mathcal{K}_\nu}{\mathcal{K}^2}
 \biggr)
 \rho^{\mu\nu}_\rmii{V,NS}(\mathcal{K}) 
 \nn &  \equiv &
  \int_\mathcal{X} 
   e^{i \mathcal{K}\cdot \mathcal{X}}
  \left\langle
    \fr12 \bigl[ 
    \hat{\mathcal{J}}^\mu (\mathcal{X}), 
    \hat{\mathcal{J}}^{ }_{\mu}(0)
    \bigr]
  \right\rangle^{ }_{\rmi{c}}
  \;, \quad
    \hat{\mathcal{J}}^\mu \equiv  \hat{\bar{\psi}}\gamma^\mu\hat{\psi}
  \;, \la{rho_NS}
\ea 
where $c$ denotes a connected quark contraction.  

The spectral function can be represented as an imaginary part 
of a retarded correlator which, in turn, is an analytic 
continuation of an imaginary-time (Euclidean) correlator: 
\be
 \rho^{ }_\rmii{NS}(\mathcal{K}) \; = \;
 \im \Pi^{ }_\rmii{R}(\mathcal{K}) \; =  \;
 \im \left. \Pi_\rmii{E} \right|^{ }_{k_n \to - i [\ko + i 0^+]}
 \;. \la{cut0}
\ee
The imaginary-time correlator is defined as 
\be
 \Pi^{ }_\rmii{E}(K) \equiv 
 \int_0^{1/T} \!\! {\rm d}\tau \! \int_\vec{x} 
 e^{i K\cdot X}
 \, 
 \Bigl\langle
   ( \bar{\psi}\gamma^\mu \psi )(\tau,\vec{x}) \,
   ( \bar{\psi}\gamma_\mu \psi )(0,\vec{0})
 \Bigr\rangle^{ }_T
 \;, \la{PiE}
\ee
where $K \equiv (k_n,\vec{k})$, with $k_n = 2\pi n T$, $n\in\mathbbm{Z}$
denoting bosonic Matsubara frequencies; 
$K\cdot X = k_n \tau - \vec{k}\cdot\vec{x}$; 
and  
$
 \langle ... \rangle^{ }_T
$
denoting a thermal expectation value. 
We first compute the imaginary-time correlator, and then determine
the spectral function from \eq\nr{cut0}.

Before proceeding let us briefly elaborate on a more general case, 
with the imaginary-time correlator
\be
 \Pi^{\mu\nu}_\rmii{E}(K) \equiv 
 \int_0^{1/T} \!\! {\rm d}\tau \! \int_\vec{x} 
 e^{i K\cdot X}
 \, 
 \Bigl\langle
   ( \bar{\psi}\gamma^\mu \psi )(\tau,\vec{x}) \,
   ( \bar{\psi}\gamma^\nu \psi )(0,\vec{0})
 \Bigr\rangle^{ }_T
 \;. \la{munu_E}
\ee
Because of current conservation all 
components are not independent; at finite temperature there are two 
independent structures. In terms of the spectral function, we can write
\be
 \im \Pi^{\mu\nu}_{\rmii{R}} (\mathcal{K})
 \; = \;
  \mathbbm{P}_\rmii{T}^{\mu\nu} (\mathcal{K}) 
   \, \rho^{ }_\rmii{T}(\mathcal{K}) + 
   \mathbbm{P}_\rmii{L}^{\mu\nu} (\mathcal{K}) 
   \, \rho^{ }_\rmii{L}(\mathcal{K})
 \;, \la{munu}
\ee
where the projectors can be defined as 
\ba
 \mathbbm{P}_\rmii{T}^{\mu\nu} (\mathcal{K}) & = & 
 - {\eta^{\mu }}_{ i} {\eta^{\nu }}_{ j} 
   \biggl(\delta^{ }_{ij} - \frac{k_i k_j}{k^2} \biggr)
 = {\eta^{\mu }}_{ i} {\eta^{\nu }}_{ j} 
   \biggl(\eta^{ }_{ij} + \frac{\mathcal{K}_i \mathcal{K}_j}{k^2} \biggr)
 \;, \la{PiT} \\ 
 \mathbbm{P}_\rmii{L}^{\mu\nu} (\mathcal{K}) & = & 
 \eta_{ }^{\mu\nu} - \frac{\mathcal{K}^\mu \mathcal{K}^\nu}{\mathcal{K}^2}
 - \mathbbm{P}_\rmii{T}^{\mu\nu} (\mathcal{K})
 \;. \la{PiL}
\ea
Here the metric convention $\eta_{\mu\nu} \equiv \mbox{diag}(+---)$
is assumed, and latin indices correspond to spatial directions. 
The mode labelled L is longitudinal with respect to 
the three-momentum $\vec{k}$; 
however, it is transverse with respect to $\mathcal{K}$.
It is seen from \eq\nr{rho_NS} that 
\be
 \rho^{ }_\rmii{NS}(\mathcal{K}) \; = \; 
 \im \Pi^{ }_\rmii{R}(\mathcal{K})\; = \;
 { \bigl\{ \im \Pi^\rmii{ }_\rmii{R}(\mathcal{K}) \bigr\}^\mu }_\mu \; = \; 
 2 \rho^{ }_\rmii{T}(\mathcal{K}) + \rho^{ }_\rmii{L}(\mathcal{K})
 \;, \la{NS_T_L}
\ee
and most of our discussion concerns this combination. 
However, in connection with lattice data 
in \se\ref{se:eucl} the two structures $\rho^{ }_\rmii{T}$ and  
$\rho^{ }_\rmii{L}$ are addressed separately.

The imaginary-time correlator of \eq\nr{PiE}
can be computed with regular path integral
techniques. If evaluated perturbatively, 
its expression can be ``scalarized'', or reduced to a sum of 
a few independent ``master'' sum-integrals, all of which have an 
O(4) invariant appearance. 
Taking subsequently the cut defined in \eq\nr{cut0}, which 
removes terms independent of $K$, and 
choosing to work in dimensional regularization, with   
$D = 4 - 2\epsilon$ denoting the space-time dimension,  
the NLO expression reads
\ba
 \im \Pi_\rmii{R} & = & 4(1-\epsilon)\Nc 
 \; \rho^{ }_{\mathcal{J}^{ }_\rmii{b}}
 \nn & + & 
 8 (1-\epsilon) g^2 \Nc \CF \; \Bigl\{ 
  2 \bigl[ 
      \rho^{ }_{\mathcal{I}^{ }_\rmii{b}}
     - \rho^{ }_{\bar{\bar{\mathcal{I}}}^{ }_\rmii{b}}
    \bigr]
 + 2(1-\epsilon) \bigl[ 
      \rho^{ }_{\mathcal{I}^{ }_\rmii{d}}
     - \rho^{ }_{\bar{\bar{\mathcal{I}}}^{ }_\rmii{d}}
    \bigr]
 \nn & & \; + \,  
 2 \epsilon \, 
 \rho^{ }_{\mathcal{I}^{ }_\rmii{f}}
 - 
 \frac{3+2\epsilon}{2} \,
 \rho^{ }_{\mathcal{I}^{ }_\rmii{g}}
 + 
 2(1+\epsilon)
 \rho^{ }_{\mathcal{I}^{ }_\rmii{h}}
 + 
 2(1-\epsilon)
 \rho^{ }_{\mathcal{I}^{ }_\rmii{h'}}
 - 
 \rho^{ }_{\mathcal{I}^{ }_\rmii{j}}
 \Bigr\} + \rmO(g^4) 
 \;. \la{imPiE}
\ea
The coupling $g^2 \equiv 4 \pi \alpha_s$ is the renormalized one, 
and $\Nc = 3$, $\CF \equiv (\Nc^2 - 1)/(2\Nc)$.
Apart from all possible $2\leftrightarrow 2$ scatterings, the NLO 
corrections incorporate $1\leftrightarrow 3$ scatterings as well as
virtual corrections to $1\leftrightarrow 2$ scatterings~\cite{master}.
The individual master spectral functions in \eq\nr{imPiE} stand for
\be
 \rho^{ }_{\mathcal{I}^{ }_\rmii{x}} \equiv
 \im [ \mathcal{I}^{ }_\rmii{x} ]^{ }_{k^{ }_n \to -i [\ko + i 0^+]} 
 \;, \la{cut}
\ee
where the labelling refers to a notation
employed in refs.~\cite{bulk_wdep,shear_wdep} 
(the definitions needed are repeated in appendix~A). 
The statistics of the different propagators are 
identified by indices $\sigma_0, ..., \sigma_5$
as illustrated in \eq\nr{labelling}; more specifically, 
the two combinations appearing in \eq\nr{imPiE} 
carry the statistics
\ba
 \rho^{ }_{\mathcal{I}^{ }_\rmii{x}} 
  & \Leftrightarrow &
 (\sigma_0\, \sigma_1\, \sigma_2\, \sigma_3\, \sigma_4\, \sigma_5 ) = (+----+)
 \;,  \la{index1} \\
 \rho^{ }_{\bar{\bar{\mathcal{I}}}^{ }_\rmii{x}} 
  & \Leftrightarrow &
 (\sigma_0\, \sigma_1\, \sigma_2\, \sigma_3\, \sigma_4\, \sigma_5 ) = (+-++--)
 \;.  \la{index2}
\ea
Details concerning the evaluation of the different 
$ \rho^{ }_{\mathcal{I}^{ }_\rmii{x}} $'s 
can be found in refs.~\cite{master,rel}. 

Each of the master spectral functions can be written as
\be
 \rho^{ }_{\mathcal{I}^{ }_\rmi{x}} 
 = 
  \rho^{\rmi{vac}}_{\mathcal{I}^{ }_\rmi{x}} 
 + 
  \rho^{\rmii{$T$}}_{\mathcal{I}^{ }_\rmi{x}} 
 \;, \la{split_up}
\ee
where $  \rho^{\rmi{vac}}_{\mathcal{I}^{ }_\rmi{x}} $ 
denotes a vacuum part.
Only the vacuum parts have divergences at NLO; therefore, 
in coefficients multiplying the thermal parts, we can 
set $\epsilon \to 0$. The results obtained after these substitutions
are given in the next section.

%
\section{Main results}
\la{se:results}

%
\subsection{Strict NLO expression}

Rewriting \eq\nr{imPiE} after the splitup in \eq\nr{split_up},
a subsequent expansion in $\epsilon$ up to $\rmO(\epsilon^0)$, 
and the insertion of known vacuum terms 
as listed in ref.~\cite{rel}
(the vacuum term of $  \rho^\rmii{ }_{\mathcal{I}^{ }_\rmii{j}} $
from ref.~\cite{master} vanishes), we get
\ba
 \im \Pi_\rmii{R} & = & - \frac{ \Nc T \mathcal{K}^2 }{2\pi k }
 \; \biggl\{ 1 + \frac{3 \alpha_s \CF}{4\pi} \biggr\} \; 
 \ln\biggl\{
   \frac{\cosh\bigl(\frac{\kp}{2 T} \bigr) }
        {\cosh\bigl(\frac{\km}{2 T} \bigr) }  
 \biggr\}
 \nonumber\\[2mm] & + & 
 32 \pi \alpha_s \Nc \CF \; \Bigl\{ 
  2 \bigl[ 
      \rho^\rmii{$T$}_{\mathcal{I}^{ }_\rmii{b}}
     - \rho^\rmii{$T$}_{\bar{\bar{\mathcal{I}}}^{ }_\rmii{b}}
     + \rho^\rmii{$T$}_{\mathcal{I}^{ }_\rmii{d}}
     - \rho^\rmii{$T$}_{\bar{\bar{\mathcal{I}}}^{ }_\rmii{d}}
    \bigr]
 \; - \,  
 \frac{3}{2} \,
 \rho^\rmii{$T$}_{\mathcal{I}^{ }_\rmii{g}}
 + 
 2 \bigl[ 
 \rho^\rmii{$T$}_{\mathcal{I}^{ }_\rmii{h}}
 + 
 \rho^\rmii{$T$}_{\mathcal{I}^{ }_\rmii{h'}}
 \bigr]
 - 
 \rho^\rmii{$T$}_{\mathcal{I}^{ }_\rmii{j}}
 \Bigr\}
 \nonumber\\[2mm] &  + & \rmO(\alpha_s^2) 
 \;. \la{imPiE_NLO}
\ea
Here the light-cone momenta $\mbox{\hspace*{5cm}}$
\be
 k^{ }_\pm \equiv \frac{\ko \pm k}{2} > 0 
 \la{kpm}
\ee 
have been defined. In addition we denote  
\be
 M \equiv \sqrt{\mathcal{K}^2} > 0 \; ; 
\ee
this ``photon mass'' is real 
in the time-like domain considered. 

In \fig\ref{fig:NLO}, 
the LO and NLO results for $- \im \Pi_\rmii{R}/T^2$ are plotted
as a function of $M/T$ for various values of $k/T$. 
Only the time-like domain relevant for \eq\nr{physics} is shown. 
It can be observed that for 
$M \gsim \pi T$, any (Lorentz violating) 
dependence on the spatial momentum $k/T$ is modest. 
For $M \ll \pi T$ the loop expansion breaks down because
the NLO term overtakes the LO term. 
(This regime is discussed in more detail in \se\ref{ss:soft}.)  

\begin{figure}[t]


\centerline{%
 \epsfysize=7.5cm\epsfbox{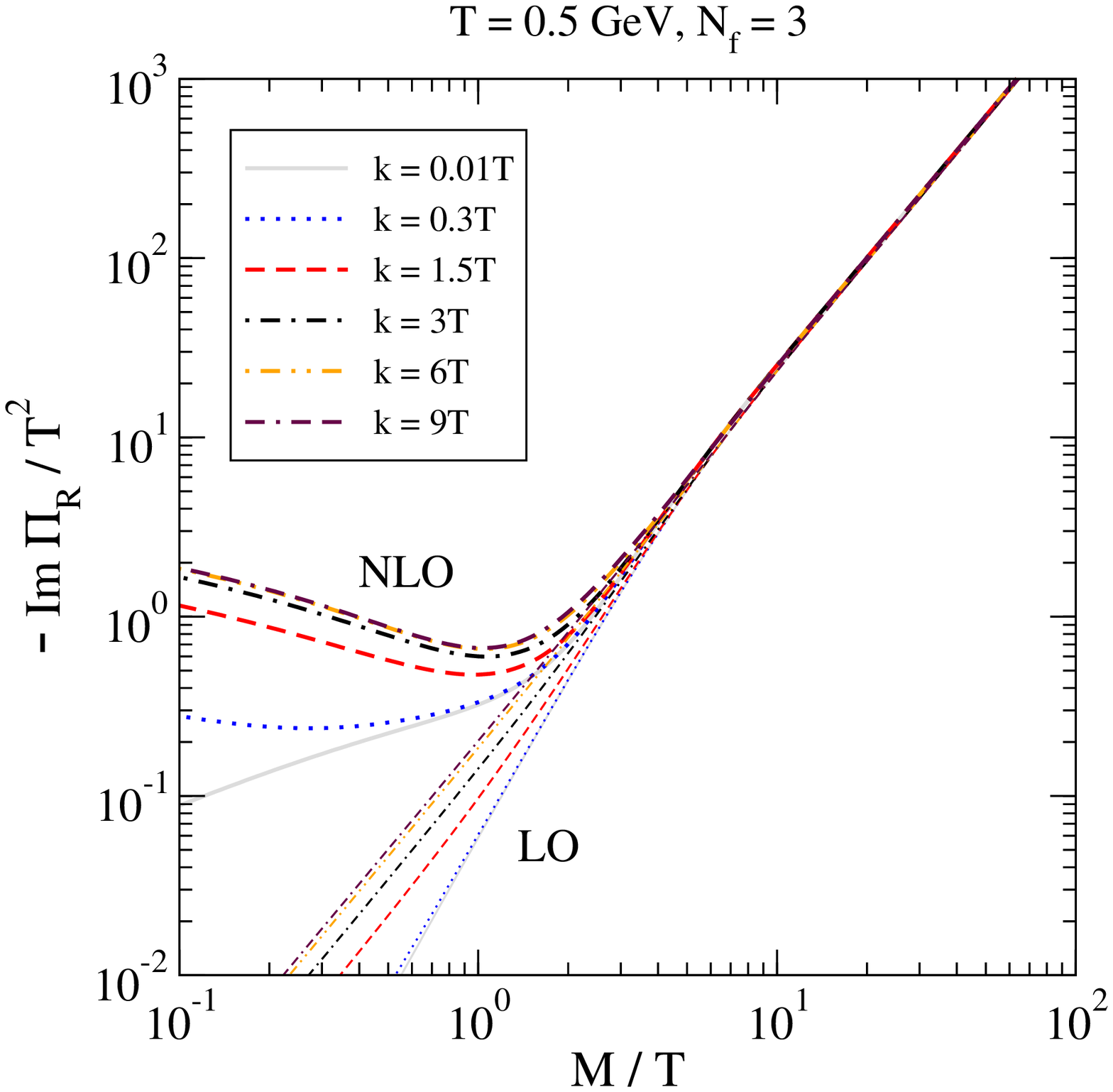}%
~~~\epsfysize=7.5cm\epsfbox{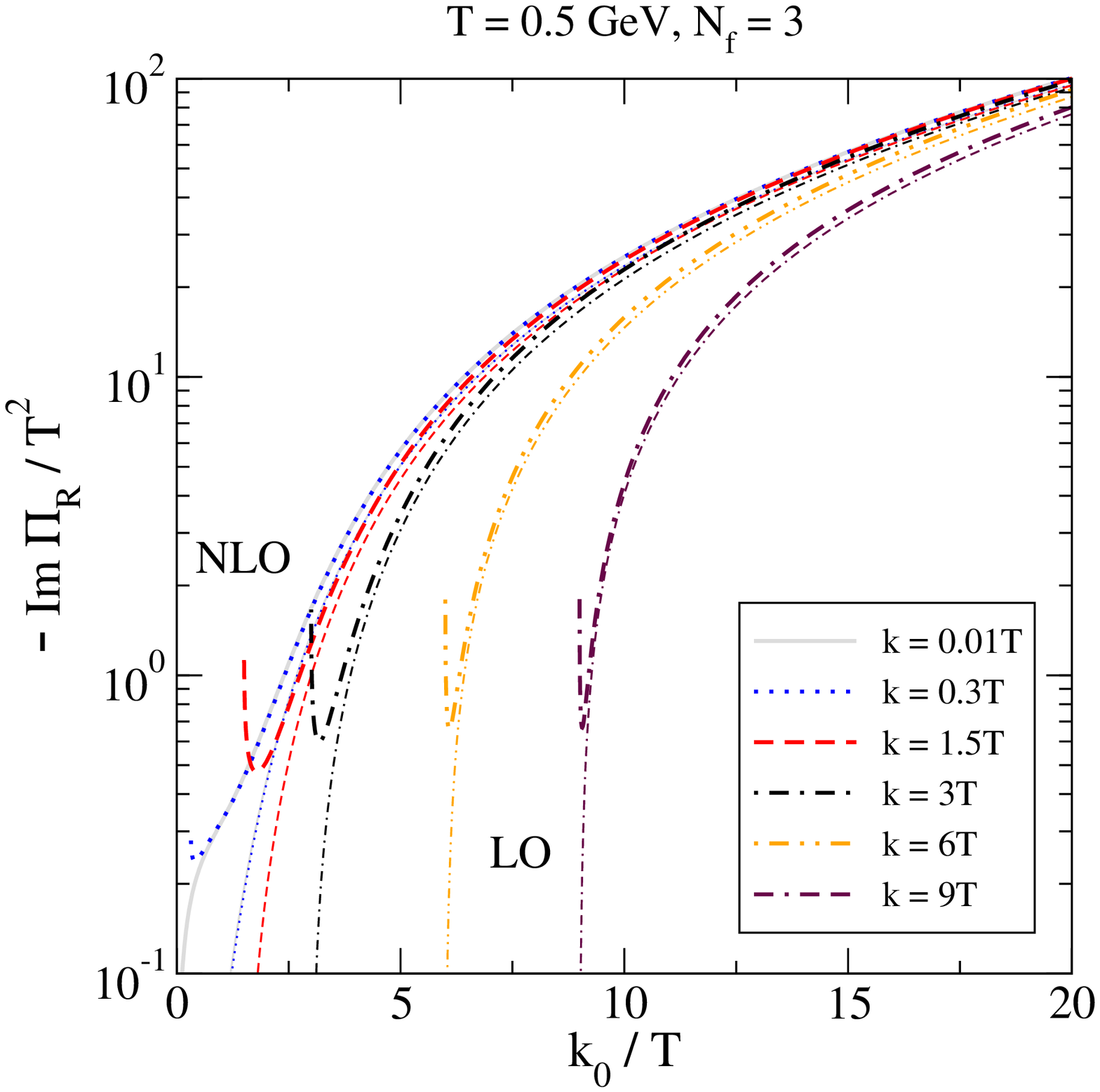}
}

\caption[a]{\small
Strict loop expansion up to NLO. The same data is shown in two ways, 
as a function of $M/T$ (left) and as a function of $\ko / T$ (right),
with $\ko \ge \sqrt{k^2 + (0.1T)^2}$ in the latter case.  
The gauge coupling and the renormalization scale have been 
fixed as specified in appendix~B ($\bmu = \bmu^{ }_\rmi{ref}$ here).
}

\la{fig:NLO}
\end{figure}

%
\subsection{Hard limit}

The result of \eq\nr{imPiE_NLO} can be simplified in a ``hard'' 
limit $M \gg \pi T $, in which Operator Product
Expansion (OPE) techniques become available~\cite{sch}. 
In fact each of the master spectral functions can be expanded
separately~\cite{bulk_ope,shear_ope,nonrel}, displaying 
an expansion of the form 
\be
 \rho^{ }_{\mathcal{I}^{ }_\rmii{x}} 
 \sim M^2 + T^2 + \frac{T^4}{M^2} + 
 \rmO\Bigl( \frac{T^6}{M^4} \Bigr)
 \;. \la{ope}
\ee
When summed 
together, terms of $\rmO(T^2)$ cancel~\cite{sch}.
The remaining expression reads 
\be
 - \im\Pi^{ }_\rmii{R} 
 = \frac{\Nc {M}^2}{4\pi}
 \; \biggl\{ 1 + \frac{3 \alpha_s \CF}{4\pi} \biggr\} \; 
 + 
 \frac{16 \alpha_s \Nc \CF}{3}
 \frac{\ko^2 + k^2/3}{M^4}
 \int_p p (4 \nF{} - \nB{})
 + 
 \rmO\Bigl( \frac{\alpha_s T^6}{M^4} \Bigr)
 \;, \la{imPiE_ASY}
\ee 
where $\nF{}$ is the Fermi distribution and 
\be
 \int_{p} p\, \nB{} = \frac{\pi^2T^4}{30}
 \;, \quad
 \int_{p} p\, \nF{} = \frac{7\pi^2T^4}{240}
 \;. \la{nBnF}
\ee

In \fig\ref{fig:ASY} the expression from \eq\nr{imPiE_ASY}
is compared with the full result from \eq\nr{imPiE_NLO}. 
It is observed that the OPE results are accurate
for $M \gsim 8 T $. This is somewhat sooner than 
for generic individual NLO master spectral functions~\cite{master,rel}; 
the reason is that the LO result has only 
exponentially small thermal corrections for $M \gg \pi T $, 
so that large power corrections 
appearing in the full result
are suppressed by $\rmO(\alpha_s)$. 

\begin{figure}[t]


\centerline{%
 \epsfysize=7.5cm\epsfbox{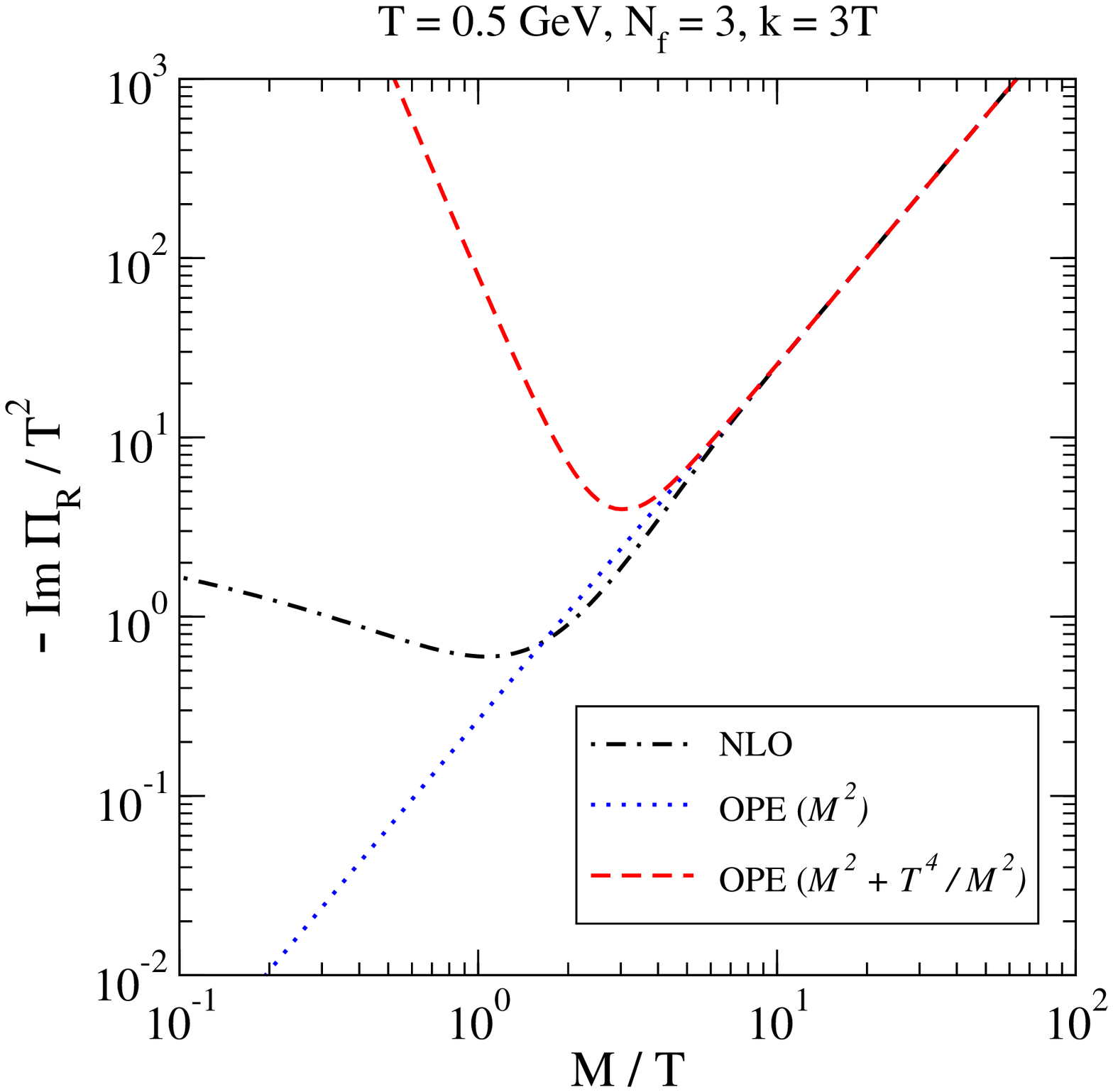}%
~~~\epsfysize=7.5cm\epsfbox{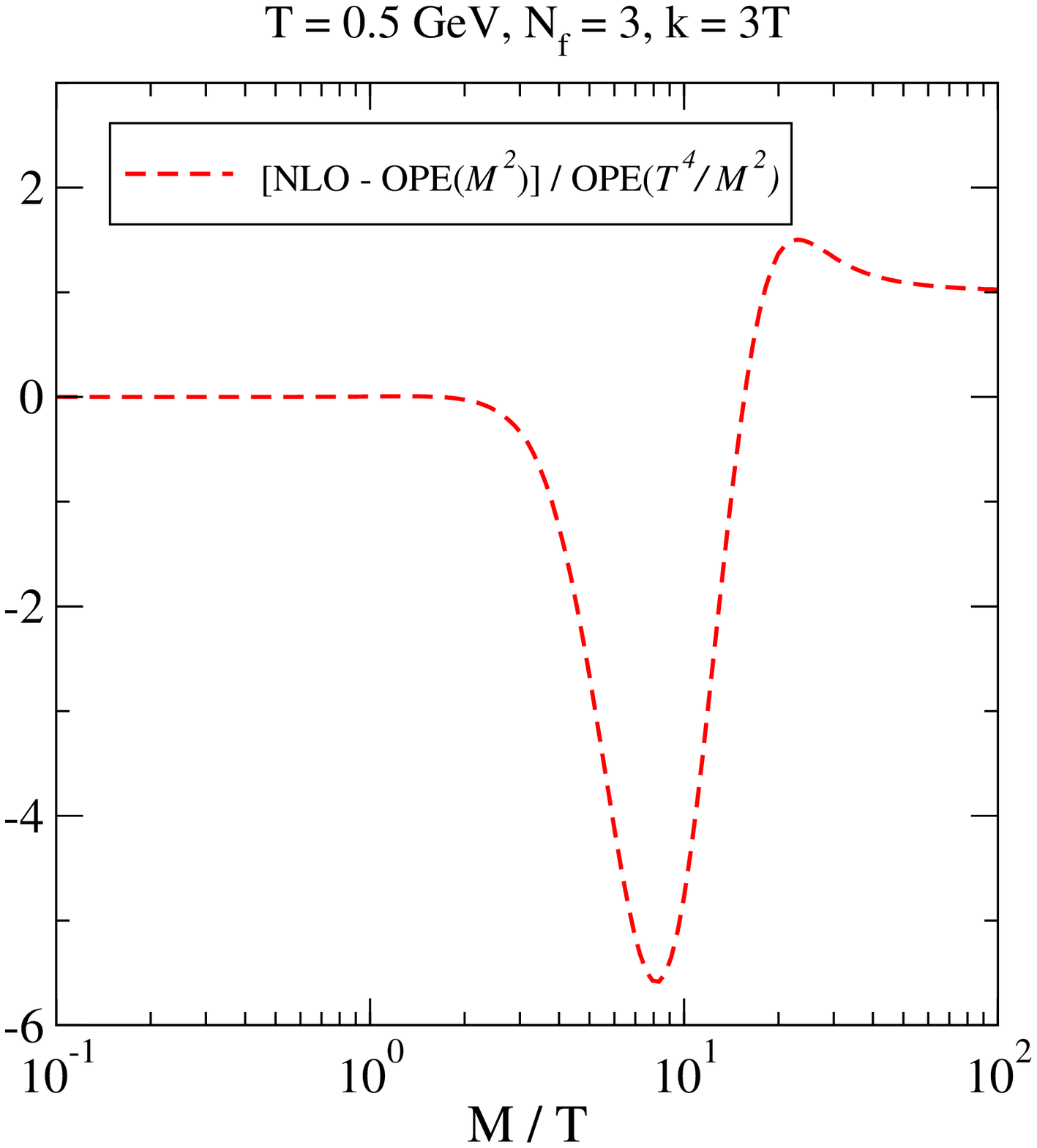}
}

\caption[a]{\small
Left: Comparison of the NLO result from 
\eq\nr{imPiE_NLO} at $k=3T$ with the OPE formula from 
\eq\nr{imPiE_ASY}, the latter evaluated up to various orders as indicated in 
the parentheses. Right: A relative difference probing 
the convergence of the expansion, with unity indicating that the 
$\rmO(T^4/M^2)$ 
correction dominates the remainder.  
}

\la{fig:ASY}
\end{figure}

%
\subsection{Towards the soft limit}
\la{ss:soft}

As is visible in \fig\ref{fig:NLO}, the NLO correction overtakes
the LO term when $\ko \to k^+$, and therefore the loop expansion breaks 
down. In this regime infinitely many loop orders need to be resummed in order
to obtain a consistent weak-coupling result. The technique goes under
the name of the Landau-Pomeranchuk-Migdal (LPM) resummation
(the Hard Thermal Loop (HTL) resummation is an ingredient but 
not sufficient on its own), and has been 
implemented for $k\sim \pi T$ in ref.~\cite{agz} and 
for $k = 0$ in ref.~\cite{mr}. The outcome cannot be expressed in analytic
form, but requires a numerical solution of an inhomogeneous Schr\"odinger-type
equation with a light-cone potential describing interactions. 

Even though we do not discuss the soft regime $|\ko - k| \ll \pi T$
systematically in the present paper, there are some qualitative remarks
that can be made. Traditionally, one feature assigned to the soft regime
is the generation of thermal masses to otherwise massless particles; 
in particular, for energetic
quarks with $k \gsim \pi T$ the concept of an ``asymptotic'' thermal 
mass, denoted by 
\be
 m_\infty^2 = \frac{g^2 \CF T^2}{4} 
 \;,
\ee
is assumed to be relevant~\cite{gT}. Recomputing {\em naively}
with the mass included,  
the LO result for the time-like domain reads~\cite{aarts}
(this is often referred to as a thermal Drell-Yan process)
\be
 \im \Pi_\rmii{R}^\rmii{LO,$m_\infty$} =
 -\frac{\Nc ({M}^2 + 2 m_\infty^2) T}{2\pi k}
   \ln\left[
   \frac{
         \cosh\Bigl(\frac{\ko + k \sqrt{1 - 4 m_\infty^2/{M}^2}
                           }{4 T} \Bigr) }
        {
         \cosh\Bigl(\frac{\ko - k \sqrt{1 - 4 m_\infty^2/{M}^2}
                           }{4 T} \Bigr) }  
   \right] 
   \, \theta \bigl( 
   \ko - \sqrt{k^2 + 4 m_\infty^2} \bigr)
   \;.  \la{imPiE_TREE}
\ee
Now, if $m_\infty^2 \ll M^2$, as is the case in the regime in which 
our computation is valid, \eq\nr{imPiE_TREE} can be expanded to first
non-trivial order in $m_\infty^2$: 
\be
 \im \Pi_\rmii{R}^\rmii{LO,$m_\infty$}  = 
 -\frac{\Nc({M}^2 + 2 m_\infty^2) T}{2\pi k}
  \ln\biggl\{
   \frac{\cosh\bigl(\frac{\kp}{2 T} \bigr) }
        {\cosh\bigl(\frac{\km}{2 T} \bigr) }  
 \biggr\}
 + \frac{\Nc m_\infty^2}{2\pi}
 \bigl[
   1 - \nF{}(\kp) - \nF{}(\km) 
 \bigr]
 + \rmO(m_\infty^4)
 \;. \la{mass_resum}
\ee
Remarkably, 
it can be verified that the $\rmO(m_\infty^2)$-terms here 
match {\em exactly} the contributions of the master spectral functions
$
      \rho^\rmii{$T$}_{\mathcal{I}^{ }_\rmii{b}}
$,
$
     \rho^\rmii{$T$}_{\bar{\bar{\mathcal{I}}}^{ }_\rmii{b}}
$,
$
     \rho^\rmii{$T$}_{\mathcal{I}^{ }_\rmii{d}}$
and 
$
     \rho^\rmii{$T$}_{\bar{\bar{\mathcal{I}}}^{ }_\rmii{d}}
$
in \eq\nr{imPiE_NLO}, cf.\ \eqs(B.15) and (B.22) of ref.~\cite{rel}.  
These master spectral functions are special in that they are the only
ones containing a factorized thermal tadpole integral. 
Therefore, naive thermal mass resummation can be ``topologically''
justified through the NLO computation. In contrast, it does not have
a power-counting justification: 
even though the factorized master spectral functions
overtake the LO result at $M \ll \pi T$ (the LO
result vanishes whereas these structures remain
finite), these are not the dominant terms
at $M \ll \pi T$, cf.\ \fig\ref{fig:Is}. 

The dominant master spectral function at $M \ll \pi T$
is the one denoted by $\rho^\rmii{$T$}_{\mathcal{I}^{ }_\rmi{h'}}$
(cf.\ \eq\nr{def_Ihp}), which 
diverges logarithmically in this limit:
\be
 \rho^\rmii{$T$}_{\mathcal{I}^{ }_\rmi{h'}} 
  \stackrel{M \,\ll\, \pi T}{\approx}
  -\frac{1}{32\pi} \ln \Bigl( \frac{T^2}{M^2} \Bigr)
  \Bigl[ 1 - 2 \nF{}(k) \Bigr]
  \int_p \frac{\nB{}(p) + \nF{}(p)}{p}
 \;, \la{Ihp_ir}
\ee
where 
$
 \int_p \nB{}/p = T^2/12
$, 
$
 \int_p \nF{}/p = T^2/24
$.
This well-known divergence~\cite{pvr} 
(the factor $-2\nF{}(k)$ is often omitted)
can be traced back to $2\leftrightarrow 2$ 
scatterings with soft momentum exchange; the phase space distributions
originate from the familiar structures~\cite{simple} 
of gain and loss terms\footnote{%
 The loss terms are eliminated if the additional factor $\nB{}(k)$
 from \eq\nr{physics} is multiplied in. 
 }
of a Boltzmann equation 
[only quarks ($q$) and gluons ($g$) have phase space distributions; 
photons ($\gamma$) are not part of the medium]:
\ba
 &   & 
 \nF{}(k)\, \nB{}(p)\, [1 - \nF{}(p)]\; - \; 
  [1-\nF{}(k)]\, [1+\nB{}(p)]\, \nF{}(p)
 \hspace*{12mm}  \mbox{$q(k)\, g(p) \leftrightarrow \gamma(k)\, q(p)$}
 \hspace*{4mm}   \nn 
  & + & 
 \nF{}(k)\, \nF{}(p)\, [1 + \nB{}(p)]\; - \;
 [1-\nF{}(k)]\, [1-\nF{}(p)]\, \nB{}(p)
 \hspace*{12mm} \mbox{$q(k)\, \bar{q}(p) \leftrightarrow \gamma(k)\, g(p)$}
 \hspace*{4mm} \nn 
 & = & 
 -  [ 1 - 2 \nF{}(k)] \, [\nB{}(p) + \nF{}(p)]
 \;.
\ea 
The divergence is lifted by Landau damping of the exchanged
nearly-static quarks~\cite{kapusta,baier} (cf.\ ref.~\cite{cg}
for an overview), an effect 
that becomes visible after an HTL resummation. In any case 
\eq\nr{Ihp_ir} is not among the effects of simple
mass resummation, \eq\nr{mass_resum}. (Even after the 
actual divergence has been lifted, the loop expansion still 
breaks down, because the NLO term overtakes the LO term
for $ M \lsim g T$.)

To summarize, the most significant NLO corrections at
$M \ll \pi T$ are not related to thermal mass generation, 
which has therefore {\em not} been implemented in the current study. 
They result rather from soft scatterings, and are 
as such a precursor to the enhancement
over the LO result that has been found previously through more 
complete computations in this corner of the $(k,\ko)$-plane.
A strength of the current ``straightforward'' 
analysis is that subtle 
issues of double-counting that have plagued resummed 
computations are avoided. Nevertheless, the current analysis
breaks down at the latest when $\ln(\pi T/M) \gg 1$. 

%
\section{Dilepton spectra}
\la{se:spectra}

We proceed to computing dilepton production spectra.
Going over to physical units, {\em viz.}
\be
  \frac{{\rm d} N_{\mu^-\mu^+}}
   {{\rm d}^4 \mathcal{X} {\rm d}^4 \mathcal{K}}
 \times \mbox{GeV}^4\mbox{fm}^4 
 = \frac{{\rm d} N_{\mu^-\mu^+}}
   {{\rm d}^4 \mathcal{X} {\rm d}^4 \mathcal{K}} 
   \biggl( \frac{1000}{197.327} \biggr)^4
  \;, 
\ee 
results are shown for $\Nf = 3$, 
fixing $\Lambdamsbar\simeq 360$~MeV~\cite{pacs}, 
in \fig\ref{fig:rates}. 
The renormalization scale and its variation 
are chosen as specified in appendix~B.
Two temperatures are considered, 
and at each temperature results are plotted as a function of the 
photon energy $\ko$, for fixed values of the invariant
photon mass $M$. For $T = 1$~GeV a good overall agreement 
with the results of ref.~\cite{agz} can be observed
(on a logarithmic scale),
despite the very different approximations inherent to the computations. 
For $M \gsim 1$~GeV the results of the present
study are more accurate than previous ones and, 
judging from the scale dependence, contain
uncertainties on a 10--30 percent level.\footnote{%
 The numerical results displayed in \fig\ref{fig:rates} can be downloaded from
 {www.laine.itp.unibe.ch/dilepton-nlo/}. 
 } 

\begin{figure}[t]


\centerline{%
 \epsfysize=7.5cm\epsfbox{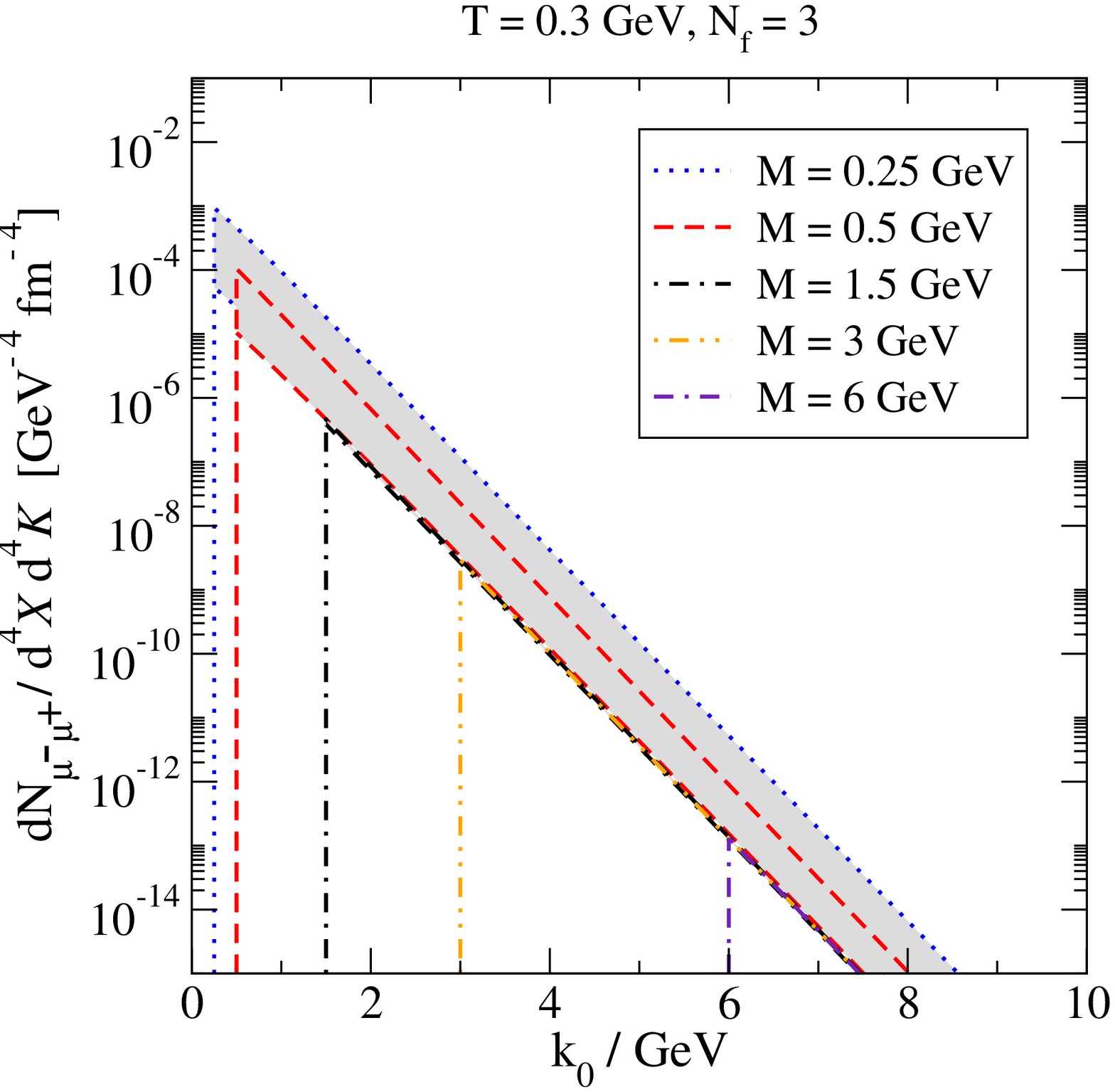}%
~~~\epsfysize=7.5cm\epsfbox{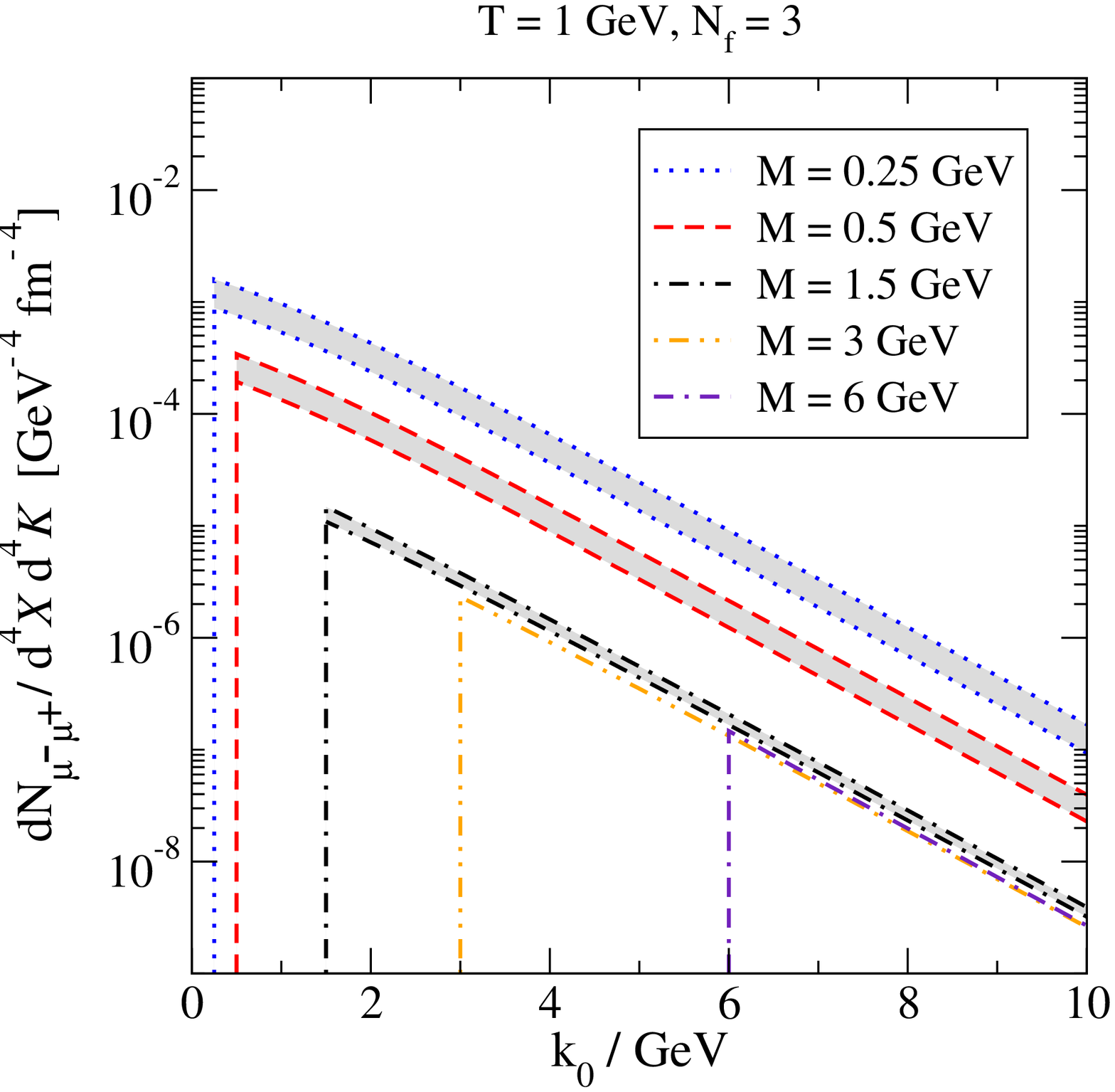}
}

\caption[a]{\small
Thermal dilepton rates according to NLO perturbation theory, 
for $T = 0.3$~GeV (left) and $T=1$~GeV (right), 
as a function of photon energy. The plots are for $\Nf = 3$
and $\Lambdamsbar = 360$~MeV~\cite{pacs}. 
Bands from scale variation are shown for the three smallest 
photon masses (cf.\ appendix~B).
The case $T = 1$~GeV is shown in order to permit for a comparison with
fig.~7 of ref.~\cite{agz}; the results are close except for an additional
spike at the smallest $\ko$ for 
$M < 1.5$~GeV in ref.~\cite{agz}. (Peculiarly it appears that
completely correct LO results for $M \ll \pi T$ have not been plotted
in the literature~\cite{cg}.)
}

\la{fig:rates}
\end{figure}

%
\section{Imaginary-time correlators}
\la{se:eucl}

%
\subsection{General considerations}

The imaginary-time correlator corresponding to a spectral function
($\rho = \im \Pi^{ }_\rmii{R}$)
antisymmetric in $\ko \to - \ko$ is given by 
\be
 G_\rmii{E}(\tau) = 
 \int_0^\infty \! \frac{{\rm d}\ko}{\pi} \, \im \Pi^{ }_\rmii{R} (\ko,\vec{k}) 
 \frac{\cosh\Bigl( \frac{1}{2T} - \tau \Bigr) \ko }
 {\sinh\Bigl(\frac{\ko}{2T} \Bigr)}
 \;. \la{relation}
\ee
A powerlike growth of $\im \Pi^{ }_\rmii{R}$ 
at $\ko \gg \pi T$ leads to a powerlike 
divergence of $ G_\rmii{E}(\tau) $ at $\tau \ll 1/T$; this divergence should
be subtracted from numerical data~\cite{analytic} in order for a well-defined
analytic continuation to be possible at least in principle~\cite{cuniberti}. 
The power divergence is determined by employing 
a vacuum spectral function 
in \eq\nr{relation}.
In the following we start
by working out ``reference'' imaginary-time correlators 
which contain a vacuum-like spectral 
function modified by LO thermal corrections (\se\ref{ss:eucl_hard}); 
subsequently issues related to NLO and higher
thermal modifications are commented upon (\se\ref{ss:eucl_soft}).  

According to \eq\nr{munu} the vector spectral function
contains two independent parts. 
In this paper, we have computed the contraction
$\im { \{\Pi^{ }_\rmii{R}
 \}^\mu}_\mu = 2 \rho^{ }_\rmii{T} + \rho^{ }_\rmii{L}$
up to NLO in the time-like domain $\ko \ge k$, 
cf.\ \eq\nr{NS_T_L}.  
There are two separate challenges
which prohibit a direct comparison with 
the continuum-extrapolated lattice measurements reported
in ref.~\cite{hengtong2}:
\bi

\item[(i)]
According to \eq\nr{relation} the space-like domain $\ko < k$ contributes 
to $G_\rmii{E}(\tau)$, even though it plays no role for
the dilepton rate, which includes a prefactor 
$\theta(\mathcal{K}^2 - 4 m_\mu^2)$ 
for a finite muon mass. {\em The space-like domain
was not worked out in the present paper at NLO.}

\item[(ii)]
In ref.~\cite{hengtong2} the momentum was chosen as $\vec{k} = (k,0,0)$
and the components $G^{ }_{11}$  and $G^{ }_{22} = G^{ }_{33}$ of the 
vector correlator were analyzed. It can be deduced from 
\eqs\nr{munu}--\nr{PiL} that these
correlators are determined by the spectral functions 
$
 {\{\im \Pi^{ }_\rmii{R}\}}_{11} = -  
 \ko^2\, \rho^{ }_\rmii{L} / \mathcal{K}^2
$ 
and 
$
 {\{\im \Pi^{ }_\rmii{R}\}}_{22} = 
 {\{\im \Pi^{ }_\rmii{R}\}}_{33} = - \rho^{ }_\rmii{T}
$,
respectively
(for the time component,  
$
 {\{\im \Pi^{ }_\rmii{R}\}}_{00} = -  
 k^2\, \rho^{ }_\rmii{L} / \mathcal{K}^2
$). 
Unfortunately knowledge of $G^{ }_{11}$ in configuration space does not 
allow us to extract $G^{ }_{00}$, 
because the Ward identity 
\be
 \partial_\tau^2 G^{ }_{00}(\tau) = k^2 G^{ }_{11}(\tau)
 \;, \quad 0 < \tau < \frac{1}{T} 
 \;, \la{Ward}
\ee
does not have a unique solution. {\em Due to a missing
$G^{ }_{00}$ the results of ref.~\cite{hengtong2} are not 
sufficient for extracting the full vector correlator
$
 G^{ }_{00} - G^{ }_{ii} 
$. }

\ei

To rectify the second problem, all that is needed is an estimate of
$G^{ }_{00}$, which is presumably simply a matter of analyzing 
existing data. To overcome the first problem, a dedicated
study of the domain $\ko < k$ is needed. 
In the next section we do consider $\ko < k$ at LO, 
but NLO corrections are left to future work. 

%
\subsection{Contribution from hard physics}
\la{ss:eucl_hard}

In order to estimate the imaginary-time correlators 
$G^{ }_{11}$  and $G^{ }_{22} = G^{ }_{33}$ measured in 
ref.~\cite{hengtong2}, information is needed about 
the two spectral functions $\rho^{ }_\rmii{T}$, 
$\rho^{ }_\rmii{L}$  appearing in \eq\nr{munu}. 
These are more complicated
than the sum $2\rho^{ }_\rmii{T} + \rho^{ }_\rmii{L}$; for instance, at LO, 
\ba
 \rho^\rmii{LO}_\rmii{T} & = & 
 -\frac{2 \Nc \mathcal{K}^2}{k^2}
 \biggl[
    \frac{\ko^2 + k^2}{2} \bigl\langle 1 \bigr\rangle 
  - 2 \bigl\langle p(\ko - p) \bigr\rangle  
 \biggr]
 \;, \la{rhoT} \\ 
 \rho^\rmii{LO}_\rmii{L} & = & 
 + \frac{4 \Nc \mathcal{K}^2}{k^2}
 \biggl[
    \frac{\ko^2 - k^2}{2} \bigl\langle 1 \bigr\rangle 
  - 2 \bigl\langle p(\ko - p) \bigr\rangle  
 \biggr]
 \;, \la{rhoL}
\ea
where 
\ba
 \langle ... \rangle 
 & \equiv &  \frac{1}{16\pi k}
 \biggl\{ 
 \theta(\km)
 \int_{\km}^{\kp} \! {\rm d}p \, 
 - 2 \theta(-\km)
 \int_{\kp}^{\infty} \! {\rm d}p \, 
 \biggr\} 
 \bigl[
   \nF{}(p - \ko) - \nF{}(p)   
 \bigr] \, (...) 
 \;. 
\ea
For $\kp > 0$ but $\km$ of either sign,
the values of the moments read (cf.\ ref.~\cite{aarts})
\ba
 \langle 1 \rangle & = & 
 \frac{1}{\pi k}
 \biggl\{
  \frac{T}{8}    \ln \biggl( 
      \frac{1 + e^{-\kp / T} } 
           {1 + e^{-|\km| / T} } 
   \biggr)
  +  \frac{\theta(\km) \, k}{16}
 \biggr\}
 \;, \la{unity} \\ 
 \langle p (\ko - p) \rangle & = & 
 \frac{1}{\pi k}
 \biggl\{
   \frac{\mathcal{K}^2 T}{32}
   \ln \biggl( 
      \frac{1 + e^{-\kp / T} } 
           {1 + e^{-|\km| / T} } 
   \biggr)
    + 
    \frac{\theta(\km)\, k(3\ko^2 - k^2)}{192}
 \nn & & \quad 
   + \,
   \frac{k T^2}{8}
   \Bigl[
     \Lii\Bigl( - e^{-\kp / T} \Bigr) 
       + \sign(\km) \, 
     \Lii\Bigl( - e^{-|\km| / T} \Bigr) 
   \Bigr]
 \nn & & \quad 
   + \, 
   \frac{T^3}{4}
   \Bigl[
     \Liii\Bigl( - e^{-\kp / T} \Bigr) 
       - 
     \Liii\Bigl( - e^{-|\km| / T} \Bigr) 
   \Bigr]
   \biggr\} 
 \;. \la{pk0p}
\ea
As is readily visible from \eqs\nr{rhoT}, \nr{rhoL}, 
the complicated  $  \langle p (\ko - p) \rangle $
drops out in \eq\nr{NS_T_L}:
\be
 { \bigl\{ \im \Pi^\rmii{LO}_\rmii{R} \bigr\}^\mu }_\mu
 = 2 \rho^\rmii{LO}_\rmii{T} + \rho^\rmii{LO}_\rmii{L} 
 = - 4 \Nc\, \mathcal{K}^2 \bigl\langle 1 \bigr\rangle
 \;. \la{tree_full}
\ee
This simplification is analogous to that 
enjoyed by the bulk channel spectral 
function, extracted from a vacuum-like Lorentz structure~\cite{bulk_wdep}, 
as compared with the shear channel one, in which a larger class of
structures  containing spatial momenta appears~\cite{shear_wdep}.

Consider now the limit 
$\kp, |\km| \gg \pi T$. Then many
terms drop out from \eqs\nr{unity}, \nr{pk0p}
and the spectral functions become
\be
 \rho^\rmii{LO}_\rmii{T} 
 \;, 
 \rho^\rmii{LO}_\rmii{L} 
 \;\;
   \stackrel{\kp,|\km|\, \gg \,\pi T}{\approx}
 \;\;
 - \frac{\Nc \mathcal{K}^2\, \theta(\mathcal{K}^2)}{12\pi }
 \;,
\ee 
so that 
\be
 \im \bigl\{ \Pi^\rmii{LO}_\rmii{R} \bigr\}^{ }_{\mu\nu}
 \;\;
   \stackrel{\kp,|\km|\, \gg \,\pi T}{\approx}
 \;\;
 \biggl( 
   \eta_{\mu\nu} - \frac{\mathcal{K}_\mu \mathcal{K}_\nu}
   {\mathcal{K}^2}
 \biggr) 
 \biggl(
 - \frac{\Nc \mathcal{K}^2}{12\pi }
 \biggr)\, \theta(\mathcal{K}^2)
 \;. 
\ee
This is simply a vacuum result. 
In vacuum, the spectral function is known 
up to 5-loop level~\cite{kit_ns,kit_si}. 
In the following we take the LO thermal 
$\rho^{ }_\rmii{T}$, $\rho^{ }_\rmii{L}$ and multiply them 
by the same {\em vacuum} factor; the results are loop-level correct
for $| \mathcal{K}^2 | \gg (\pi T)^2$ and LO correct
at $| \mathcal{K}^2 | \sim (\pi T)^2$ but in general 
underestimate thermal corrections. 

\begin{figure}[t]


\centerline{%
 \epsfysize=7.5cm\epsfbox{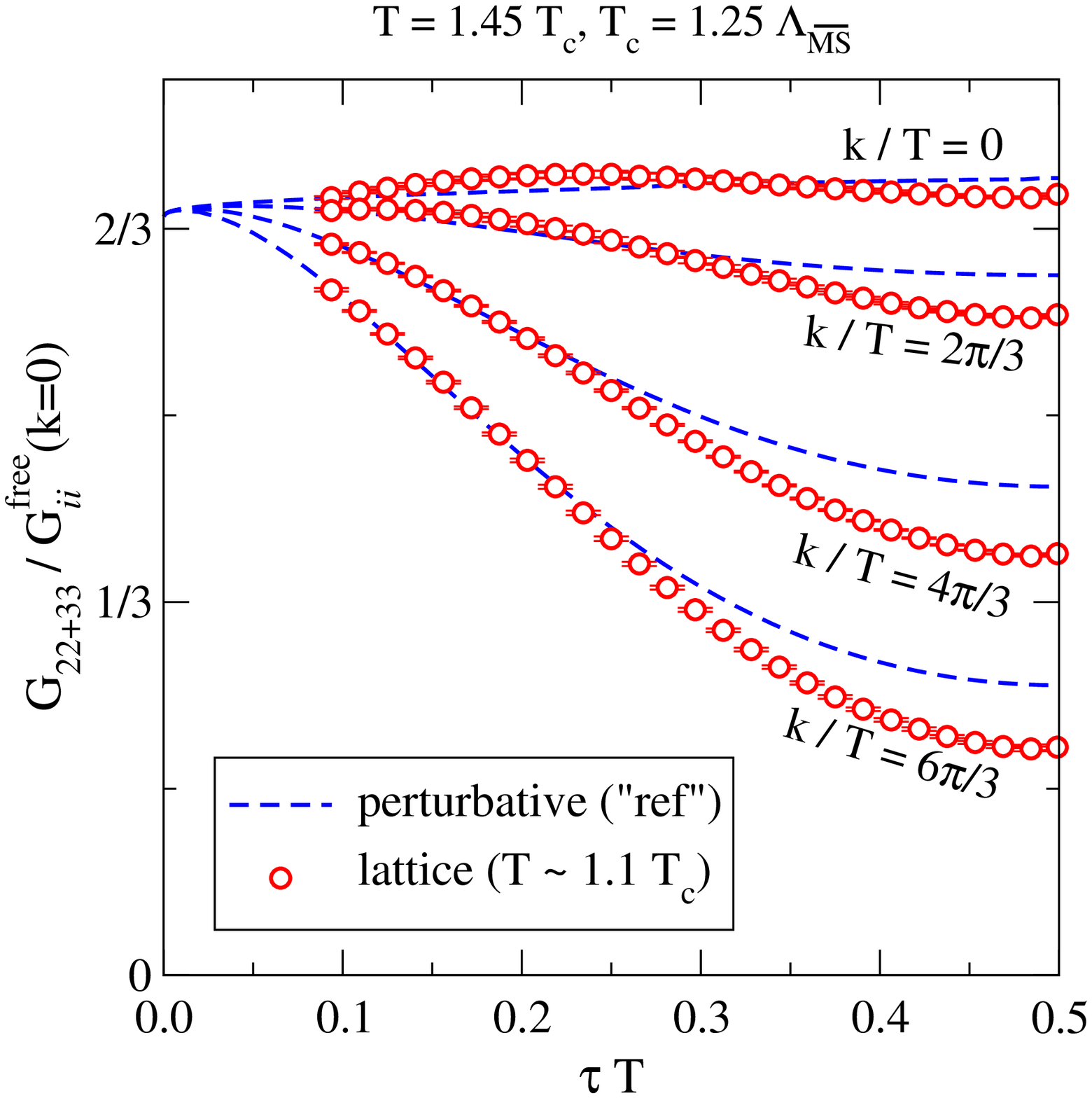}%
~~~\epsfysize=7.5cm\epsfbox{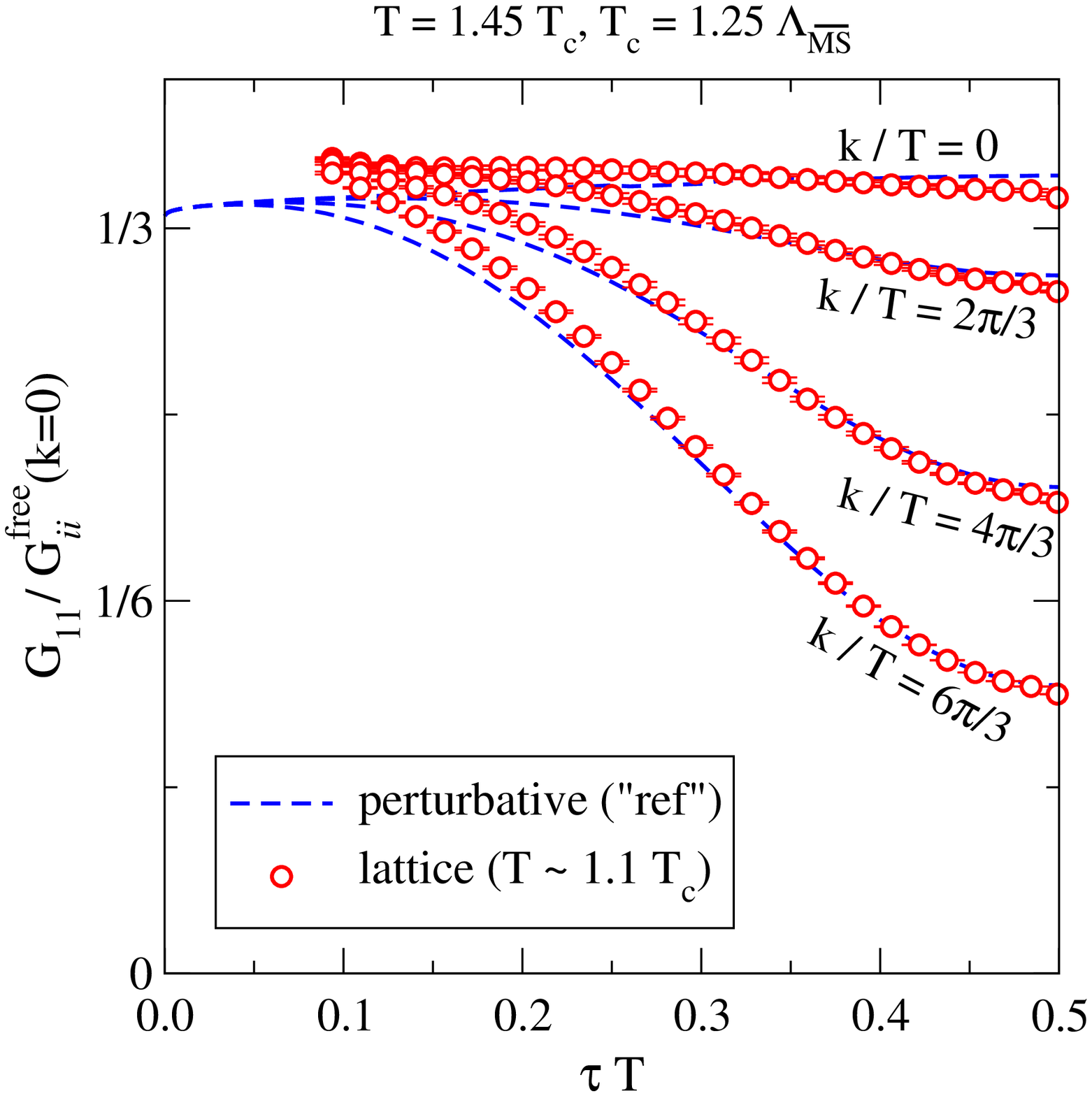}
}

\caption[a]{\small
Imaginary-time correlators based on \eq\nr{model}, 
normalized to \eq\nr{GVfree}, for $\Nf = 0$. The perturbative values are 
compared with lattice data from ref.~\cite{hengtong2}
(to make use of the data 
a continuum value of the quark-number susceptibility
is needed; we have assumed $\chi^{ }_\rmi{q} \simeq 0.88 T^2$). 
The data are for $T = 1.1\Tc$ but according to ref.~\cite{hengtong2} 
they are close to those at $T = 1.45 \Tc$ where  
the perturbative expressions were evaluated. Note that for $k=0$, 
$G^{ }_{22+33} = 2 G^{ }_{11}$, and the violation of this relation
towards small $\tau T$ is a reflection
of systematic uncertainties related to the continuum extrapolation. 
}

\la{fig:taudep_baseline}
\end{figure}

Denoting 
$
 \ell \equiv \ln({\bmu^2} / {\mathcal{K}^2})
$,
the vacuum factor~\cite{kit_ns} can be expressed as
\ba
 \mathcal{R}(\mathcal{K}^2) & \equiv &  
 \theta(\mathcal{K}^2) {\rm sign}(\ko) 
 \Bigl\{ 
  r^{ }_{0,0} + 
  r^{ }_{1,0}\, a_s + 
 \bigl( r^{ }_{2,0} + r^{ }_{2,1}\, \ell \bigr)\, a_s^2 
   \nn & + & 
 \bigl( r^{ }_{3,0} + r^{ }_{3,1}\, \ell + r^{ }_{3,2}\, \ell^2 \bigr)\, a_s^3 
    +  
 \bigl( r^{ }_{4,0} + r^{ }_{4,1}\, \ell + r^{ }_{4,2}\, \ell^2
  + r^{ }_{4,3}\, \ell^3\bigr)\, a_s^4
  + \rmO(a_s^5)
 \Bigr\}
 \;, \hspace*{8mm} \la{5l_vac}
\ea
where $a_s\equiv \alpha_s/\pi$ and the coefficients are identical 
to those listed in ref.~\cite{cond} 
(the terms 
$ r^{ }_{0,0} + 
  r^{ }_{1,0}\, a_s $
reproduce the factor 
$ 1 + \frac{3 \alpha_s \CF}{4\pi} $
from \eq\nr{imPiE_ASY}). We set
\be
 \rho_\rmii{T}^\rmi{ref} \equiv 
 \rho_\rmii{T}^\rmii{LO} \; \mathcal{R}(
   {\rm max}\{ \mathcal{K}^2, (\pi T)^2 \} )
 \;, \quad
 \rho_\rmii{L}^\rmi{ref} \equiv
 \rho_\rmii{L}^\rmii{LO} \; \mathcal{R}( 
   {\rm max}\{ \mathcal{K}^2, (\pi T)^2 \} )
 \;, \la{model}
\ee
freezing the $\mathcal{R}$-factor when entering the thermal domain.
The renormalization scale is fixed as specified in appendix~B.
The results are normalized
to the free correlator for $k=0$~\cite{ff}, 
\ba
 G_{{ii},k=0}^\rmi{free}(\tau) & \equiv & 
 6 T^3 \biggl[ 
 \pi (1-2\tau T) \frac{1 + \cos^2(2\pi\tau T)}{\sin^3(2\pi \tau T)}
  + 
 \frac{2 \cos(2\pi \tau T)}{\sin^2(2\pi \tau T)}
  + \fr16 \biggr]
 \;. \la{GVfree}
\ea
The results are shown in \fig\ref{fig:taudep_baseline}, and 
indicate good overall agreement.

%
\subsection{On soft corrections}
\la{ss:eucl_soft}

In order to be more precise than \eq\nr{model}, 
NLO {\em thermal} corrections to 
$\rho^{ }_\rmii{T}$, $\rho^{ }_\rmii{L}$
are needed for $|\mathcal{K}^2| \gsim (\pi T)^2$ and, 
perhaps more importantly, 
the soft domain $|\mathcal{K}^2| \ll (\pi T)^2$ needs to be 
properly addressed. 
If the correlators $G^{ }_{00}$, $G^{ }_{11}$, $G^{ }_{22+33}$
are inspected separately, we may expect qualitatively different 
corrections from the 
soft domain than for the 
vector channel correlator $G^{ }_{00} - G^{ }_{ii}$. One way to 
see this is that at LO both the spectral function 
$
 {\{\im \Pi^{ }_\rmii{R}\}}_{00} = -  
 k^2\, \rho^{ }_\rmii{L} / \mathcal{K}^2
$
and
$
 {\{\im \Pi^{ }_\rmii{R}\}}_{11} = -  
 \ko^2\, \rho^{ }_\rmii{L} / \mathcal{K}^2
$ 
are discontinuous across $\ko = k$,  
whereas their difference 
$
 {\{\im \Pi^{ }_\rmii{R}\}}_{00} - 
 {\{\im \Pi^{ }_\rmii{R}\}}_{11} = \rho^{ }_\rmii{L}
$
is continuous and vanishes for $\ko = k$, 
as is the case also with  
$ 
 {\{\im \Pi^{ }_\rmii{R}\}}_{22} = 
 {\{\im \Pi^{ }_\rmii{R}\}}_{33} =  
 - \rho^{ }_\rmii{T}
$. 
Therefore at LO
the light-cone regime makes a smaller contribution
to $G^{ }_{00} - G^{ }_{ii}$ than to 
$G^{ }_{00}$ and $G^{ }_{11}$.
On the other hand, after accounting for thermal loop 
corrections, $ \rho^{ }_\rmii{T} $ no longer vanishes 
at the light-cone, so NLO  
corrections may be relatively speaking
larger in $G^{ }_{22+33}$
than in $G^{ }_{00}$ and $G^{ }_{11}$.  
Indeed, 
it can be observed in \fig\ref{fig:taudep_baseline} 
that there is a larger discrepancy in $G^{ }_{22+33}$
than in $G^{ }_{11}$. 

For $k=0$, in which case there are no ambiguities 
(there is no contribution from $\ko < k$ and $G^{ }_{00}$
is a known constant),  
the discrepancy between lattice data and 
the perturbative correlator is small. 
Yet it was found in ref.~\cite{cond} that the discrepancy can be 
used for constraining the parameters of a transport peak, 
such as the diffusion coefficient $D$, 
in a non-trivial way. It will be 
interesting to see how large a discrepancy remains there for $k\neq 0$
in a continuum-extrapolated $G^{ }_{00} - G^{ }_{ii}$ and whether
it can be accounted for by the physics of the soft regime
in a similar way. As a first step strict NLO expressions, i.e.\ results from
the current paper supplemented by a similar analysis at $\ko < k$,
may be used; the logarithmic divergence they contain
at $\mathcal{K}^2 = 0$ is integrable. Going beyond this, 
LPM-resummed results could be tested, however for the 
moment none seem to exist for $\ko < k$. 
The range $\ko < k$ has been studied 
in a leading-logarithmic
approximation in ref.~\cite{ht} and within a holographic 
framework in ref.~\cite{susy}, in both cases even 
for  $\rho^{ }_\rmii{T}$ and $\rho^{ }_\rmii{L}$ separately. 
For $\ko,k\to 0$, 
$\rho^{ }_\rmii{T}$ and $\rho^{ }_\rmii{L}$
can also be parametrised by $D$ and second order transport coefficients, 
cf.\ e.g.\ ref.~\cite{ht}. 
(A general discussion of various domains can be found, for $k=0$, 
in ref.~\cite{ay}.)

%
\section{Conclusions}
\la{se:concl}
 
The purpose of this paper has been to determine
the vector channel spectral function (cf.\ \fig\ref{fig:NLO})
and the dilepton production rate (cf.\ \fig\ref{fig:rates})
up to next-to-leading order in thermal QCD, keeping track 
of a non-zero momentum of the dilepton pair with respect to 
the heat bath. The results are reliable in a characteristically 
``thermal'' regime, $\mathcal{K}^2 \gsim\, (\pi T)^2$, which for 
heavy ion collision experiments corresponds to 
$\mathcal{K}^2 \gsim\, 1~\mbox{GeV}^2$. 

In the regime considered, the expressions obtained are complicated
enough that no analytic representations have been found for all the 
structures appearing; 
we have rather
evaluated 2-dimensional integrals numerically for the different
``basis functions'' needed (cf.\ \fig\ref{fig:Is}). However, 
in a ``hard'' limit $\mathcal{K}^2 \gg (\pi T)^2$ an explicit
expression, given in 
\eq\nr{imPiE_ASY}, can be obtained~\cite{sch}. 
By comparing with the numerical 
evaluation of \fig\ref{fig:NLO}, the asymptotic
form of \eq\nr{imPiE_ASY} is seen
to be accurate in the range $\mathcal{K}^2 \gsim\, (8 T)^2$ 
(cf.\ \fig\ref{fig:ASY}).

In the opposite regime $0 < \mathcal{K}^2 \ll (\pi T)^2$, the naive
loop expansion breaks down and resummations are needed for obtaining
formally consistent results. We have observed, however, that as long as
$\ln \{ (\pi T)^2 / \mathcal{K}^2 \}$ is not large, even the naive results
of the present paper agree relatively well with resummed ones~\cite{agz}. 
The explanation could be that the resummation is 
numerically not overwhelmingly important for moderate 
$\mathcal{K}^2 \sim (gT)^2$; 
those parts of it already included in the current
expression, together with all ``hard'' processes such as $2\leftrightarrow 2$
scatterings, may capture much of the answer. 
(Resummation is also tricky in that it is non-trivial 
to avoid double counting when the resummed result 
is combined with hard processes~\cite{cg}.) 

In the regime $\mathcal{K}^2 \gsim\, (\mbox{1 GeV})^2$, 
the uncertainties of the current results could be on a 10--30\%
level, judging from the scale dependence in \fig\ref{fig:rates}.
For a comparison
with actual data, the results should be embedded in a 
hydrodynamical model incorporating the temperature evolution
of the system, which unfortunately 
goes beyond the scope of the present study.  

Apart from heavy ion data, we have elaborated on possibilities to confront
spectral functions with continuum-extrapolated lattice results
(cf.\ \fig\ref{fig:taudep_baseline}). 
This comparison is ambiguous for the moment, given that 
continuum-extrapolated data are only available for spatial
components of the vector correlator and that also 
the domain below the light-cone ($\mathcal{K}^2 < 0$) 
contributes to imaginary-time 
correlators measured on the lattice. Once these issues have 
been addressed, it appears that accounting for 
the difference of a ``hard'' perturbative
contribution and a continuum-extrapolated lattice correlator 
may permit for a non-trivial crosscheck of the 
physics of the soft domain. Apart from soft dilepton spectra, 
this might give another handle on 
the diffusion coefficient $D$, complementing
its direct estimate as a transport coefficient from
measurements at $k=0$. 

%
\section*{Acknowledgements}

I am grateful 
to A.~Francis for providing numerical data from ref.~\cite{hengtong2}
and for helpful discussions, and 
to H.B.\ Meyer and G.D.\ Moore for helpful discussions. 
This work was partly supported by the Swiss National Science Foundation
(SNF) under grant 200021-140234.

%
\appendix
\renewcommand{\thesection}{Appendix~\Alph{section}}
\renewcommand{\thesubsection}{\Alph{section}.\arabic{subsection}}
\renewcommand{\theequation}{\Alph{section}.\arabic{equation}}

%
\section{Definitions of master sum-integrals}
\la{app:A}

Denoting by $\Tinti{P}$ and $\Tinti{\{P\}}$ sum-integrals 
over bosonic and  fermionic Matsubara four-momenta, the master 
sum-integrals yielding non-vanishing cuts are defined as follows 
(a dashed line indicates a bosonic propagator, a solid line
a fermionic one, a filled blob a squared propagator, 
and a cross a structure in the numerator):
\ba
 \TopoSB(\Lsc,\Aqu,\Aqu) \quad
 \mathcal{J}^{ }_\rmi{b} & \!\!\equiv\!\! & 
 \Tint{ \{P\} } \frac{K^2}{P^2(P-K)^2}
 \;, \la{def_Jb}
 \\
%
 \TopoST(\Lsc,\Asc) \times
 \TopoSB(\Lsc,\Aqu,\Aqu) \quad
 \bar{\bar{\mathcal{I}}}_\rmi{b} & \!\!\equiv\!\! & 
 \Tint{ \{ P\} Q} \frac{1}{Q^2P^2(P-K)^2}
 \;, \la{def_Ib}
 \\
%
 \TopoST(\Lsc,\Aqu) \times
 \TopoSB(\Lsc,\Aqu,\Aqu) \quad
 {\mathcal{I}}_\rmi{b} & \!\!\equiv\!\! & 
 \Tint{\{P Q\}} \frac{1}{Q^2P^2(P-K)^2}
 \;,
 \\
%
 \TopoST(\Lsc,\Asc) \times
 \TopoSBd(\Lsc,\Aqu,\Aqu) \quad
 \bar{\bar{\mathcal{I}}}_\rmi{d} & \!\!\equiv\!\! & 
 \Tint{\{P\}Q} \frac{K^2}{Q^2P^4(P-K)^2}
 \;, \la{def_Id}
 \\
%
 \TopoST(\Lsc,\Aqu) \times
 \TopoSBd(\Lsc,\Aqu,\Aqu) \quad
 {\mathcal{I}}_\rmi{d} & \!\!\equiv\!\! & 
 \Tint{\{PQ\}} \frac{K^2}{Q^2P^4(P-K)^2}
 \;,
 \\
%
 \ToptSS(\Lsc,\Aqu,\Aqu,\Lsc) \quad
 \mathcal{I}_\rmi{f} & \!\!\equiv\!\! & 
 \lim^{ }_{\lambda \to 0}
 \Tint{\{PQ\}} \frac{1}{Q^2[(Q-P)^2 + \lambda^2](P-K)^2}
 \;, \la{def_If}
 \\
%
 \TopoSB(\Lsc,\Aqu,\Aqu) \times
 \TopoSB(\Lsc,\Aqu,\Aqu) \quad
 \mathcal{I}_\rmi{g} & \!\!\equiv\!\! & 
 \Tint{\{PQ\}} \frac{K^2}{P^2(P-K)^2Q^2(Q-K)^2}
 \;, \la{def_Ig}
 \\
%
 \ToptSAr(\Lsc,\Aqu,\Aqu,\Aqu,\Asc) \quad
 \mathcal{I}_\rmi{h} & \!\!\equiv\!\! & 
 \lim^{ }_{\lambda \to 0}
 \Tint{\{PQ\}} \frac{K^2}{Q^2P^2[(Q-P)^2+\lambda^2](P-K)^2}
 \;, \la{def_Ih}
 \\
%
 \ToptSArX(\Lsc,\Aqq,\Aqu,\Aqu,\Asc) \quad
 {\mathcal{I}}_\rmi{h'} \! & \!\!\equiv\!\! & 
 \lim^{ }_{\lambda \to 0}
 \Tint{\{PQ\}} \frac{2K\cdot Q}{Q^2P^2[(Q-P)^2+\lambda^2](P-K)^2}
 \;, \la{def_Ihp}
 \\
%
 \ToptSM(\Lsc,\Aqu,\Aqu,\Aqu,\Aqu,\Lsc) \quad
 \mathcal{I}_\rmi{j} & \!\!\equiv\!\! & 
 \lim^{ }_{\lambda \to 0}
 \Tint{\{PQ\}} \frac{K^4}{Q^2P^2[(Q-P)^2+\lambda^2](P-K)^2(Q-K)^2} 
 \;.  \la{def_Ij} \hspace*{1cm}
\ea
In order to handle different statistics simultaneously, 
a generic labelling of the lines is employed
(with individual propagators omitted for the simpler masters):
\be
 \hspace*{6mm}
 \Generic
 \hspace*{1.5cm} \;.  \la{labelling}
\ee
The labels $\sigma_0,...,\sigma_5$ take the value $+1$ for bosons 
and $-1$ for fermions. The  expressions for 
the corresponding spectral functions 
have been worked out 
in refs.~\cite{master,rel}, and we refer
to these works for more details. 
Numerical results are shown in \fig\ref{fig:Is}, 
except for the case $\mathcal{I}^{ }_\rmi{j}$ for which 
numerical results were already shown in ref.~\cite{master}.
The average momentum employed in \fig\ref{fig:Is} is defined as
\be
 k^2_\rmi{av}(M) \equiv 
 \frac{\int_0^\infty \! {\rm d}k \, k^4 
 \exp(-\frac{\sqrt{k^2 + M^2}}{T}) }
 {\int_0^\infty \! {\rm d}k \, k^2 
 \exp(-\frac{\sqrt{k^2 + M^2}}{T})} = 
 \frac{3 M T K^{ }_3(\fr{M}{T})}
    {K^{ }_2(\fr{M}{T})}
 \;. \la{kav}
\ee

\begin{figure}[t]


\centerline{%
 \epsfysize=6.4cm\epsfbox{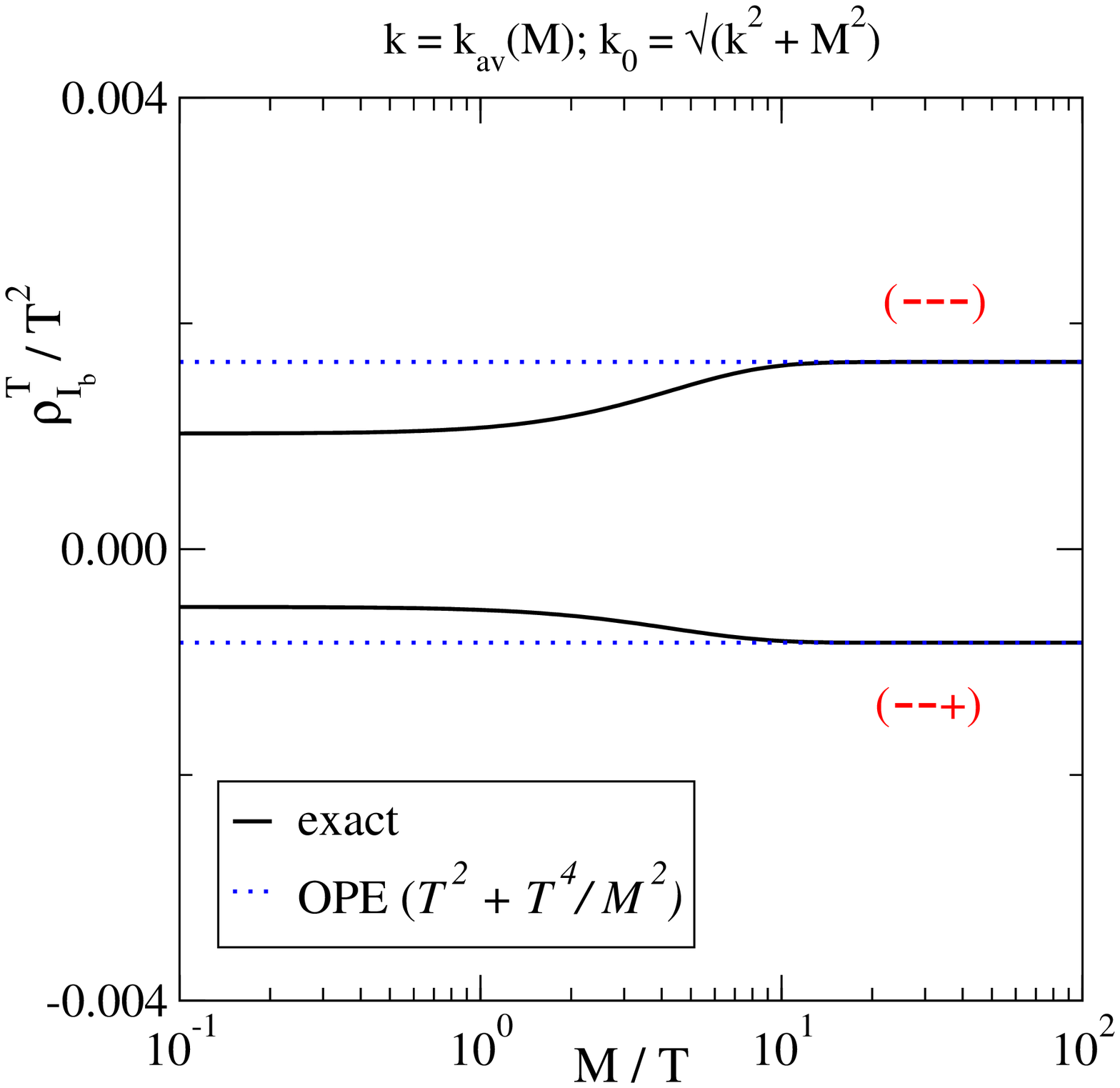}%
~~~\epsfysize=6.4cm\epsfbox{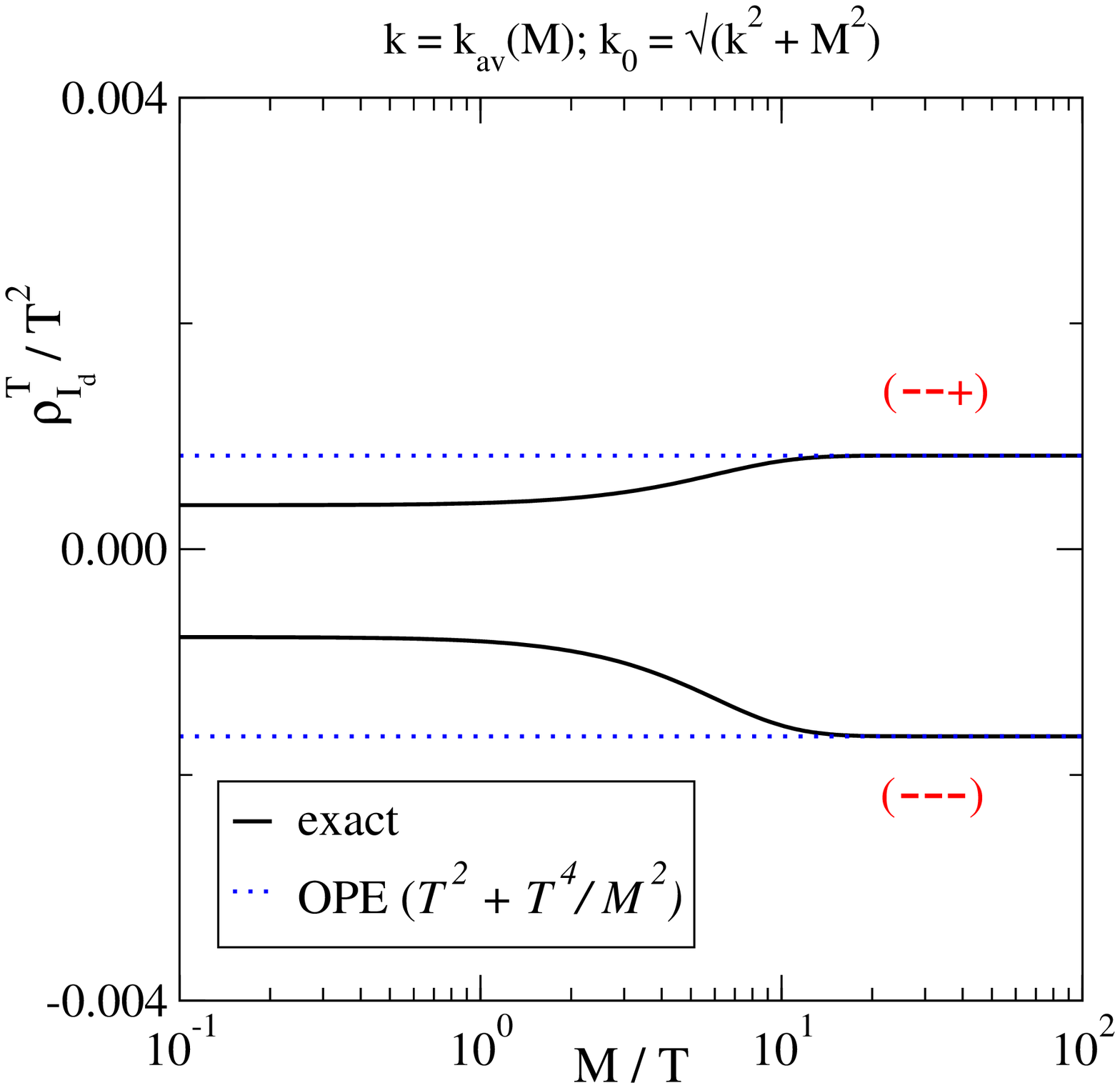}
}

\vspace*{1mm}

\centerline{%
 \epsfysize=6.4cm\epsfbox{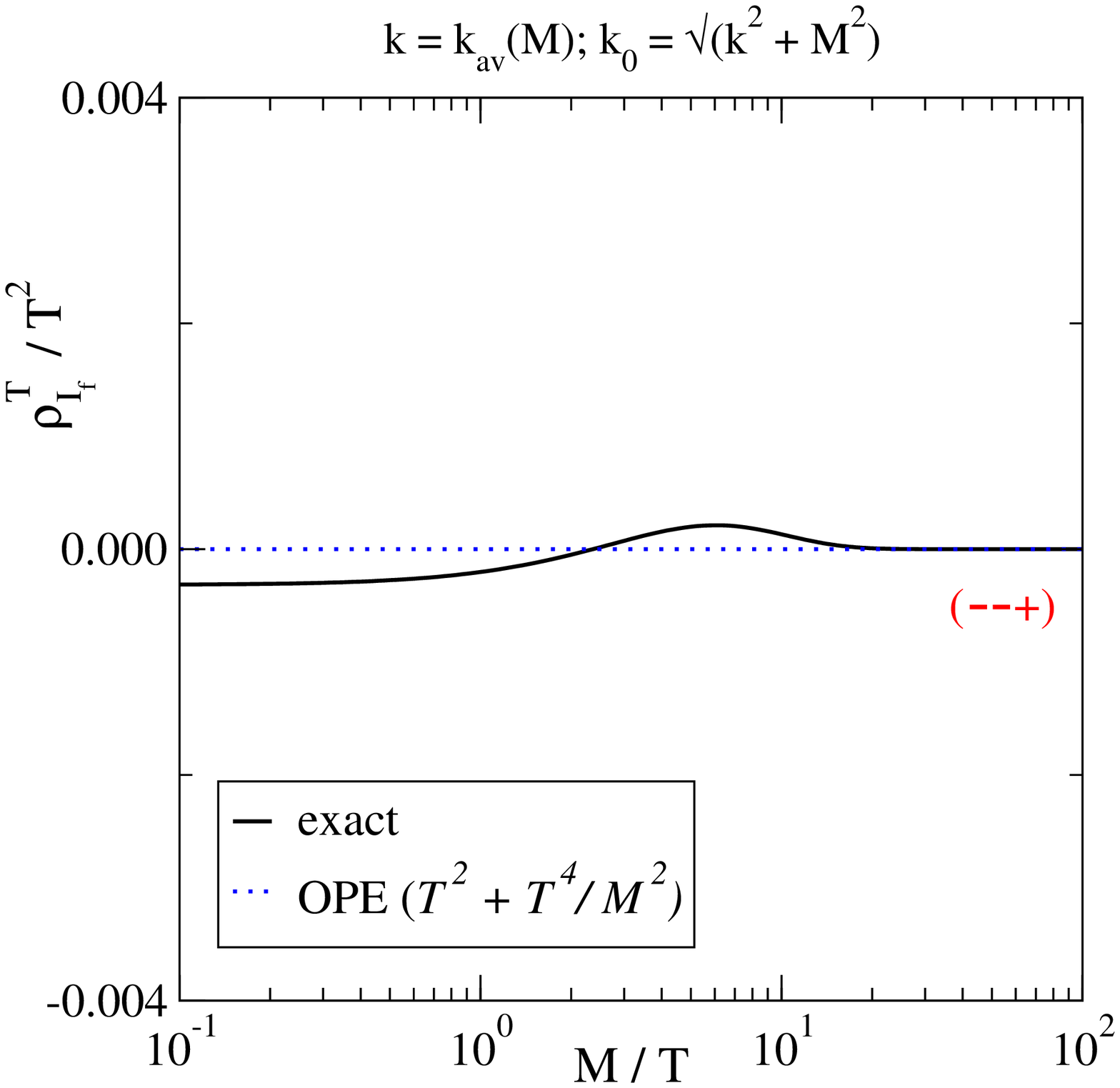}%
~~~\epsfysize=6.4cm\epsfbox{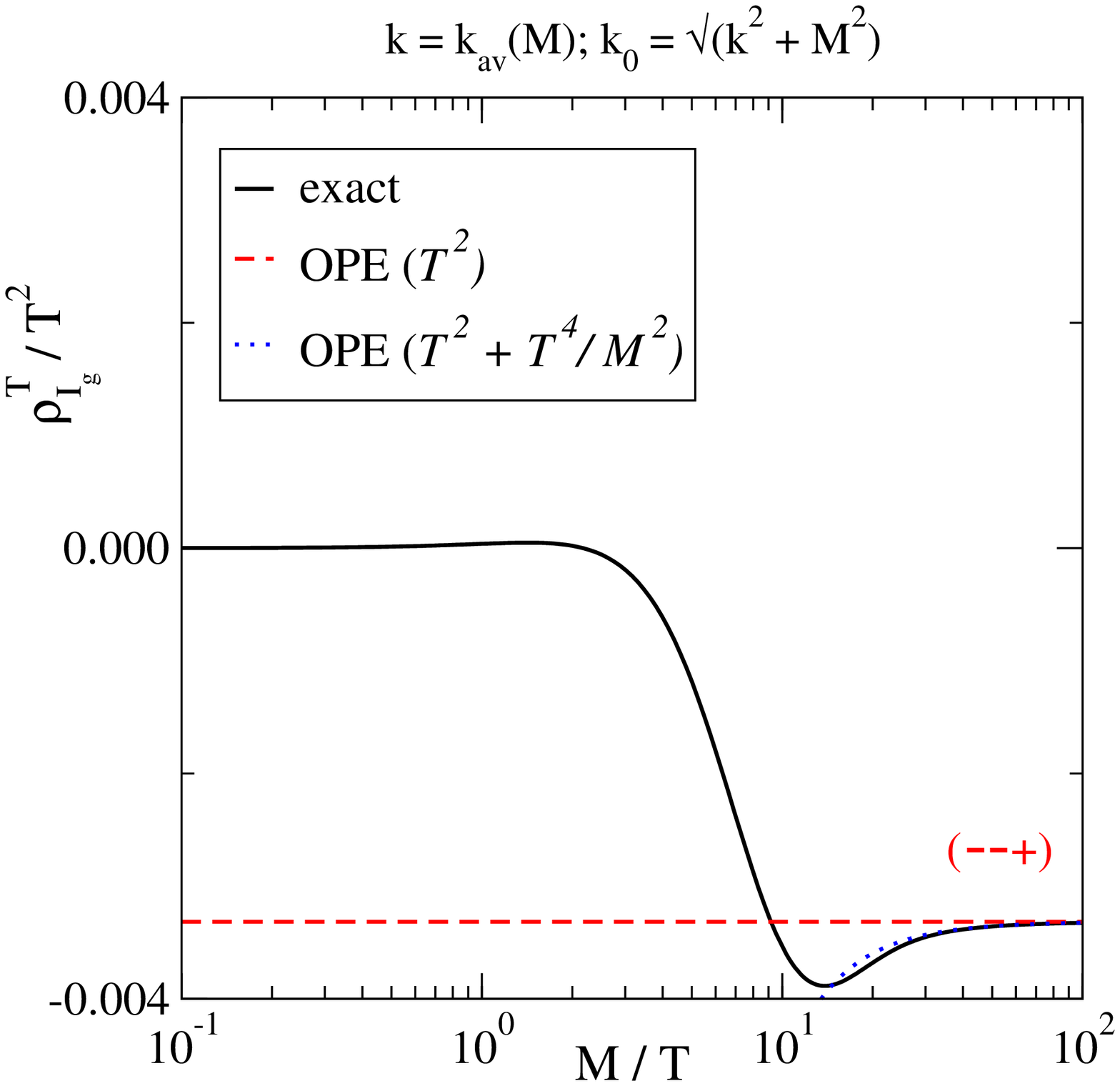}
}

\vspace*{1mm}

\centerline{%
 \epsfysize=6.4cm\epsfbox{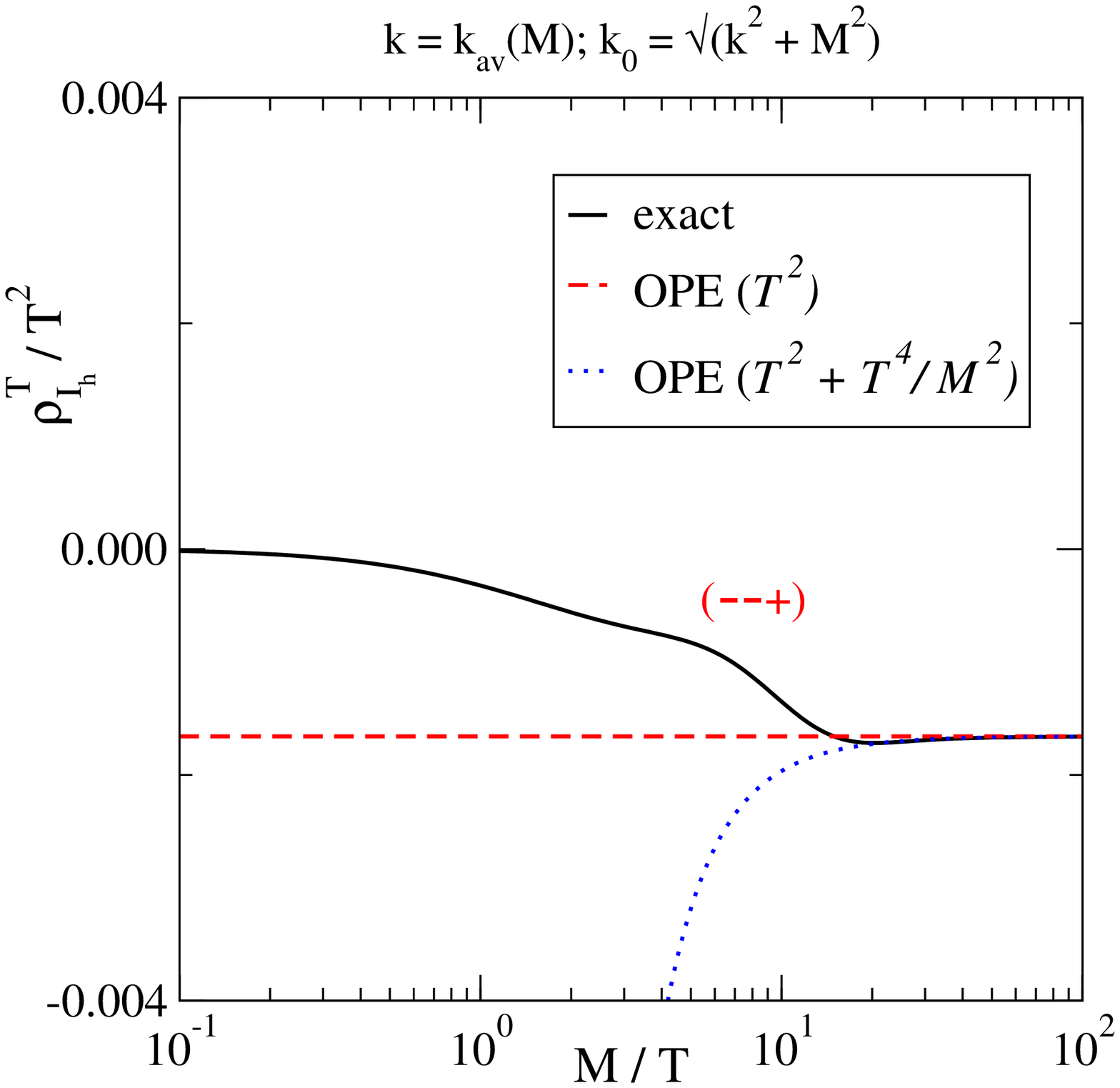}%
~~~\epsfysize=6.4cm\epsfbox{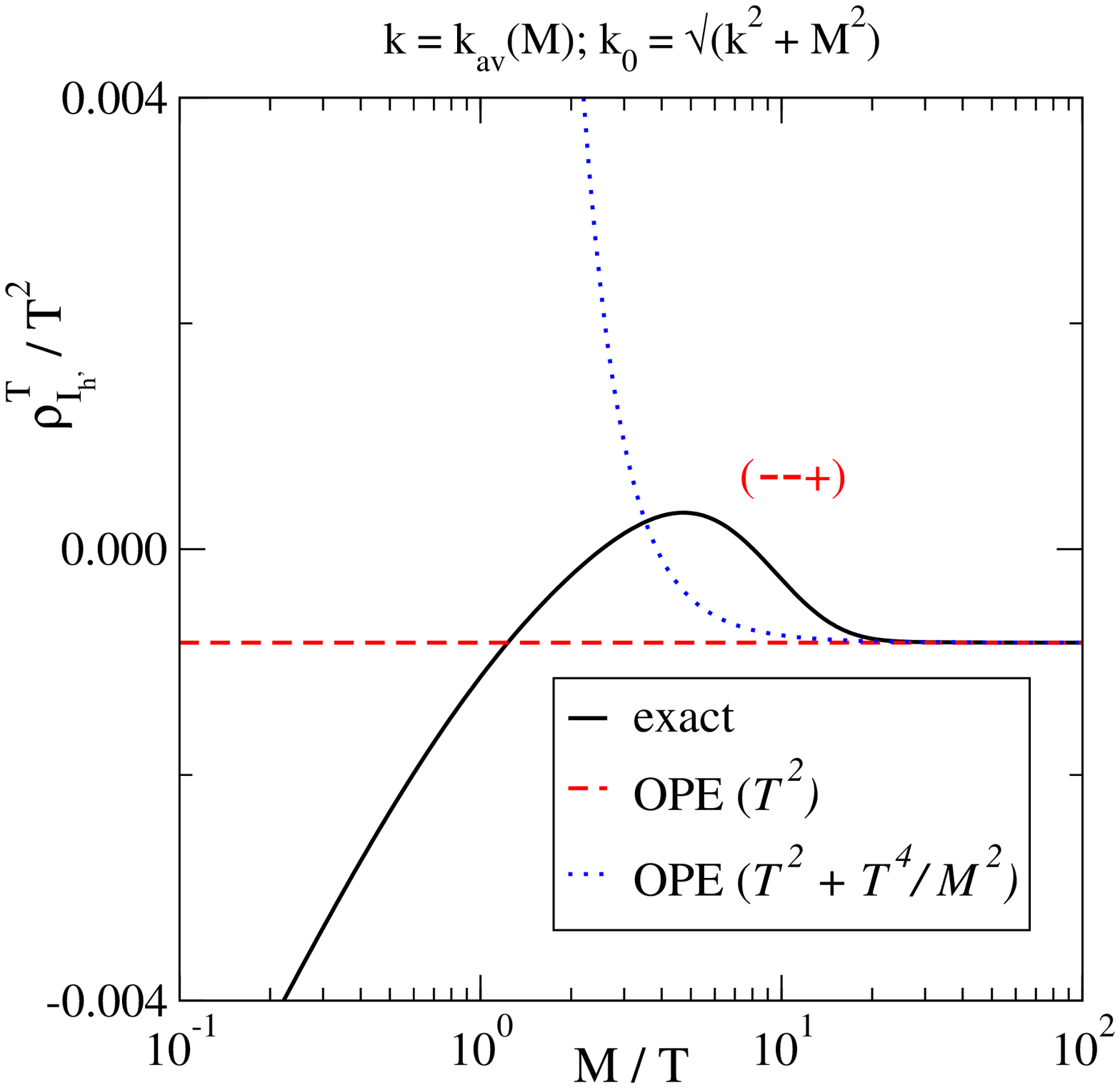}
}

\caption[a]{\small
Thermal parts of spectral functions
corresponding to \eqs\nr{def_Ib}--\nr{def_Ihp}, 
with spatial momentum fixed according to \eq\nr{kav}. The indices refer to 
$(\sigma_1 \sigma_4 \sigma_5)$, cf.\ \eqs\nr{index1}, \nr{index2}.
}

\la{fig:Is}
\end{figure}

%
\section{Choice of parameters}
\la{app:C}

The strong coupling constant runs as 
$
 \partial_t a_s = 
 - (\beta_0 a_s^2 + \beta_1 a_s^3 + \beta_2 a_s^4 + \beta_3 a_s^5 + \ldots)
$, 
where 
$
 a_s \equiv {\alpha_s(\bmu)} / {\pi}
$, 
$
 t \equiv \ln\bigl( {\bmu^2} / {\Lambda_\rmii{$\msbar$}^2} \bigr)
$, 
and, for $\Nc = 3$~\cite{rit1},  
\ba
 \beta_0 & = & \frac{11}{4} - \frac{\Nf}{6}
  \;, \quad 
 \beta_1 \; = \; \frac{51}{8} - \frac{19\Nf}{24}
  \;, \quad 
 \beta_2 \; = \; \frac{2857}{128} - \frac{ 5033\Nf}{1152}
  + \frac{325\Nf^2}{3456}
  \;, \\ 
 \beta_3 & = & \frac{149753 + 21384 \zeta(3) }{1536}
 \nn & - &
      \frac{[1078361+39048\zeta(3)]\Nf}{ 41472 } + 
      \frac{[50065+12944\zeta(3)]\Nf^2}{41472} +
      \frac{1093\Nf^3}{186624}
  \;. 
\ea
The scale parameter $\Lambdamsbar$
represents an integration 
constant and is chosen so that the asymptotic ($t \gg 1$) 
behaviour reads
\be
  a_s = \frac{1}{\beta_0 t} - \frac{\beta_1 \ln t}{\beta_0^3 t^2}
  + \frac{\beta_1^2 (\ln^2 t - \ln t - 1)  + \beta_2 \beta_0}
    {\beta_0^5 t^3} 
  + \rmO\Bigl(\frac{1}{t^4}\Bigr)
 \;.
\ee
The renormalization scale is varied within the range 
\be
 \bmu \in 
 (0.5 ... 2.0)\, \bmu_\rmi{ref}
 \;, \quad
 \bmu_\rmi{ref}^2 \equiv {\rm max}\{ \mathcal{K}^2 , (\pi T)^2  \}
 \;. \la{band}
\ee
In general we have employed 3-loop running 
(i.e.\ $\beta_0, \beta_1, \beta_2$) given that this corresponds
to the formal accuracy of \eq\nr{5l_vac}, however we have checked that
results obtained with 4-loop running are well within the error band
obtained from \eq\nr{band}. 

%

\end{document}